\documentclass[11pt,american,twoside,a4paper]{article}

\usepackage[T1]{fontenc}
\usepackage[utf8]{inputenc}
\usepackage[american,provide=*]{babel}
\usepackage{a4wide}
\usepackage{fourier,bbm,bm}
\usepackage{BOONDOX-calo}
\usepackage{amsmath,amsfonts,amsthm,amssymb}
\usepackage{color,xcolor,graphicx}
\usepackage{latexsym,mathtools,leftindex}
\usepackage[shortlabels]{enumitem}
\usepackage{xargs,xspace,needspace,xparse,easyReview}
\usepackage[active]{srcltx}
\usepackage[colorlinks=true]{hyperref}
\usepackage{fancyhdr}
\usepackage[toc]{appendix}
\usepackage[normalem]{ulem}
 
\colorlet{darkblue}{blue!50!black}
\hypersetup{
	colorlinks,%
	citecolor=darkblue,%
	filecolor=red,%
	linkcolor=darkblue,%
	urlcolor=darkblue,%
	pdfnewwindow=true,%
	pdfstartview={FitH}
}

\numberwithin{equation}{section}
\theoremstyle{plain}
\newtheorem{theorem}{Theorem}[section]
\AtBeginEnvironment{theorem}{\Needspace{5\baselineskip}}
\newcommand\bet{\begin{theorem}} \newcommand\eet{\end{theorem}}
\newtheorem{proposition}[theorem]{Proposition}
\AtBeginEnvironment{proposition}{\Needspace{5\baselineskip}}
\newcommand\bep{\begin{proposition}} \newcommand\eep{\end{proposition}}
\newtheorem{lemma}[theorem]{Lemma}
\AtBeginEnvironment{lemma}{\Needspace{5\baselineskip}}
\newcommand\bel{\begin{lemma}} \newcommand\eel{\end{lemma}}
\newtheorem{corollary}[theorem]{Corollary}
\AtBeginEnvironment{corollary}{\Needspace{5\baselineskip}}
\newcommand\bec{\begin{corollary}} \newcommand\eec{\end{corollary}}
\theoremstyle{definition}
\newtheorem{definition}[theorem]{Definition}
\AtBeginEnvironment{definition}{\Needspace{5\baselineskip}}
\newcommand\bed{\begin{definition}} \newcommand\eed{\end{definition}}
\newtheorem{remark}{Remark}[section]
\AtBeginEnvironment{remark}{\Needspace{5\baselineskip}}
\newcommand\ber{\begin{remark}} \newcommand\eer{\end{remark}}
\newtheorem{remarks}[remark]{Remarks}
\AtBeginEnvironment{remarks}{\Needspace{5\baselineskip}}
\newcommand\bers{\begin{remarks}} \newcommand\eers{\end{remarks}}
\AtBeginEnvironment{quote}{\Needspace{5\baselineskip}}

\newcommand\beq{\begin{equation}} \newcommand\eeq{\end{equation}}

\newcounter{smallarabics}
\newenvironment{arabicenumerate}
{\begin{list}{{\normalfont\textrm{(\arabic{smallarabics})}}}
		{\usecounter{smallarabics}\setlength{\itemindent}{0cm}
			\setlength{\leftmargin}{5ex}\setlength{\labelwidth}{4ex}
			\setlength{\topsep}{0.75\parsep}\setlength{\partopsep}{0ex}
			\setlength{\itemsep}{0ex}}}
	{\end{list}}
\newcommand{\ben}{\begin{arabicenumerate}} \newcommand{\een}{\end{arabicenumerate}}

\newcounter{smallroman}
\newenvironment{romanenumerate}
{\begin{list}{{\normalfont\textbf{(\roman{smallroman})}}}
		{\usecounter{smallroman}\setlength{\itemindent}{0cm}
			\setlength{\leftmargin}{5ex}\setlength{\labelwidth}{4ex}
			\setlength{\topsep}{0.75\parsep}\setlength{\partopsep}{0ex}
			\setlength{\itemsep}{0ex}}}
	{\end{list}}
\newcommand{\benr}{\begin{romanenumerate}} \newcommand{\eenr}{\end{romanenumerate}}

\AtBeginEnvironment{theorem}{\Needspace{5\baselineskip}}
\AtBeginEnvironment{proposition}{\Needspace{3\baselineskip}}
\AtBeginEnvironment{definition}{\Needspace{5\baselineskip}}
\AtBeginEnvironment{corollary}{\Needspace{5\baselineskip}}
\AtBeginEnvironment{lemma}{\Needspace{5\baselineskip}}
\AtBeginEnvironment{quote}{\Needspace{5\baselineskip}}

\NewDocumentCommand{\ASS}{mm}{\expandafter\newcommand\csname #1\endcsname{{\hyperref[#1]{\bf (#2)}}}}
\NewDocumentCommand{\preASS}{mm}{\expandafter\newcommand\csname pre#1\endcsname{{\hyperref[#1]{\bf (#2)}}}}

\newcommand\ZZ{{\mathbbm Z}}
\newcommand\QQ{{\mathbbm Q}}
\newcommand\RR{{\mathbbm R}}

\newcommand{\one}{\mathbf{1}}

\newcommand\e{\mathrm{e}}

\renewcommand{\i}{\mathrm{i}}

\newcommand\myoperator[1]{\mathop{\mathrm{#1}}}

\renewcommand\dim{\myoperator{dim}\nolimits}

\renewcommand\det{\myoperator{det}\nolimits}

\renewcommand\proof{\noindent{\bf Proof.\ }}
\renewcommand\qed{\hfill$\Box$}
\begin{document}

\title{One-Particle Resonances and Eigenstate Thermalization in Periodic Free-Fermion Lattices}

\bigskip
\author{T. Benoist$^1$ \quad V. Jak\v{s}i\'c$^2$  \quad  C.-A.  Pillet$^3$\quad A. Szczepanek$^4$\quad  C. Vogel$^5$
\\ \\
$^1$Univ Toulouse, INSA Toulouse, CNRS, IMT, Toulouse, France
\\ \\
$^{2}$Dipartimento di Matematica, Politecnico di Milano\\
piazza Leonardo da Vinci 32, 20133 Milano,  Italy
\\ \\ 
$^3$Universit\'e de Toulon, CNRS, CPT, UMR 7332, 83957 La Garde, France\\
Aix-Marseille Univ, CNRS, CPT, UMR 7332, Case 907, 13288 Marseille, France 
\\ \\
$^4$Institute of Mathematics, Jagiellonian University,\\ 
{\L}ojasiewicza 6, 30-348 Krak\'ow, Poland
\\ \\
$^5$Mathematics Institute, Ludwig-Maximilians-Universit\"at M\"unchen\\
Theresienstraße 39, 80333 Munich, Germany
}

\maketitle

\begin{abstract}
We study eigenstate fluctuations of the particle number in the 
nearest-neighbour free-fermion model on the periodic cubic lattice
$(\mathbb Z/L\mathbb Z)^d$. For every fixed dimension, every particle number,
every subvolume, and every normalized energy eigenstate, we prove a
bound of order $NL^{d-1}$ for the particle-number variance. The
corresponding eigenstate-thermalization estimate is of order
$L^{d-1}/(N\eta^2)$ for fluctuations of relative size $\eta$.  In particular,
at positive particle density the bound decays as the inverse linear size. The proof decomposes the centered particle-number operator according to
one-particle energy transfer and treats the zero- and nonzero-transfer
components separately. The nonzero-transfer part is controlled by a sharp
resonance bound, proved  from Mann's theorem on relations among roots
of unity, while the zero-transfer part is estimated by a Hilbert--Schmidt
argument. These bounds are optimal in every dimension: the maximal
nonzero-gap multiplicity has order $L^{d-1}$, and suitable many-body
eigenstates realize matching variance lower bounds and show that the
concentration bound cannot be improved in estimates uniform in the
tolerance parameter.

\end{abstract}

\clearpage
\tableofcontents

\section{Introduction}
We study macroscopic equilibrium in energy eigenstates of free fermions
on a periodic cubic lattice. The question is whether the number of
particles in a spatial region is concentrated around the value
prescribed by the uniform particle density, and how its fluctuations
depend on the particle number and system size. We ask this for every
energy eigenstate, including arbitrary superpositions within degenerate
energy eigenspaces. The rigorous formulation of macroscopic equilibrium
and its relation to eigenstate thermalization were developed in works
of Goldstein et al.~\cite{Goldstein2010a} and
Tasaki~\cite{Tasaki2016}. Concrete thermalization results for
free-fermion chains were subsequently established by Shiraishi and
Tasaki~\cite{ShiraishiTasaki2024} and by Tasaki~\cite{Tasaki2025}.

Our starting point is the work of Roos et al.~\cite{Roos2025} and
Hara and Koike~\cite{HaraKoike2026}. 

{\emergencystretch=1em
For the one-dimensional periodic free-fermion chain, Roos et
al.~\cite{Roos2025} prove an eigenstate thermalization estimate for the
particle number in a macroscopic interval. Their bound is of order $N^{-1}\log N$, where $N$ is the number of
particles, and is established under the assumption  that the chain
length $L$ is a prime number. In higher dimensions, they instead obtain
thermalization results after introducing a suitable random perturbation
of the Hamiltonian.\par}

For the nearest-neighbour free-fermion model considered here, Hara and Koike~\cite{HaraKoike2026}
prove macroscopic density equilibration in every  dimension,
starting from any pure state in the $N$-particle sector. At fixed
positive particle density, they identify the optimal ballistic
timescale $L$, where $L$ is the side length of the lattice:
on observation windows of duration $\mathcal T_L$ with
$\mathcal T_L/L\to\infty$, the fraction of nonequilibrium times tends
to zero in the thermodynamic limit, uniformly in the initial
state.\footnote{The dynamical results of Hara and Koike also motivate
work in preparation, in which we study the macroscopic evolution
of a density-step initial state at deterministic times on the
ballistic scale.}
Their argument also bounds particle-number fluctuations in every
energy eigenstate. We formulate  this bound with explicit dependence
on $N$ in Theorem~\ref{prop:HK-eigenstates}.

We follow the notation of Roos et al.~\cite{Roos2025} whenever possible
and refer to their work and~\cite{HaraKoike2026}
for  physical motivation and background.

Let $\Lambda_{L,d}=(\mathbb Z/L\mathbb Z)^d$
be the periodic cubic lattice of side length $L$, and let $H_0^{\rm fF}$ be
the nearest-neighbour free-fermion Hamiltonian on its fermionic Fock space.
For a subvolume $\Gamma\subset\Lambda_{L,d}$, we denote by $N_\Gamma$ the
operator counting the number of particles in $\Gamma$ and set
\[
\mu=\frac{|\Gamma|}{L^d}.
\]
Our main result states that, for every fixed $d\geq1$, there is a constant
$C_d$ such that every normalized eigenstate $\varphi$ of $H_0^{\rm fF}$ in
the $N$-particle sector $\mathcal H_N$ satisfies
\begin{equation}
\label{eq:intro-variance-bound}
\left\langle
\varphi,
\left(N_\Gamma-N\mu\right)^2\varphi
\right\rangle
\leq
C_dN\mu(1-\mu)L^{d-1}.
\end{equation}
For $\eta>0$, let
$P_{{\rm neq},\eta}=\boldsymbol 1_{\left\{\left|N_\Gamma/N-\mu\right|>\eta\right\}}$
be the spectral projection onto the subspace where the fraction of
particles in $\Gamma$ differs from $\mu$ by more than $\eta$. By Chebyshev's inequality,
\[
\|P_{{\rm neq},\eta}\varphi\|^2
\leq
\frac{\langle\varphi,(N_\Gamma-N\mu)^2\varphi\rangle}
{N^2\eta^2},
\]
and  \eqref{eq:intro-variance-bound} gives
\begin{equation}
\label{eq:intro-main-bound}
\left\|P_{{\rm neq},\eta}\varphi\right\|^2
\leq
\frac{C_d\mu(1-\mu)L^{d-1}}{N\eta^2}.
\end{equation}
No regularity assumption on $\Gamma$ is needed, and the bounds hold for all
$0<N<L^d$ and all side lengths $L$.

The scale in \eqref{eq:intro-variance-bound} is optimal. For every
$d\geq1$ and arbitrarily large $L$, we find a particle number
$N$, a normalized energy eigenstate $\varphi\in\mathcal H_N$, and
a subset $\Gamma\subset\Lambda_{L,d}$ such that
\[
\left\langle
\varphi,(N_\Gamma-N\mu)^2\varphi
\right\rangle
\geq c_dNL^{d-1},
\]
where $c_d>0$ depends only on the dimension. The factor $L^{d-1}/N$ in the concentration estimate
\eqref{eq:intro-main-bound} cannot be replaced by
$o(L^{d-1}/N)$ in a bound valid for all $\eta>0$. For suitable eigenstates $\varphi_L$ and tolerances
$\eta_L\asymp\sqrt{L^{d-1}/N}$,
$\|P_{{\rm neq},\eta_L}\varphi_L\|^2$ remains bounded away
from zero. These optimality results are proved in
Section~\ref{sec-higher-optimality}.

Adapting the argument of Hara and Koike's Proposition~2 to energy
eigenstates and retaining the dependence on the particle number
gives the following bound (see Theorem~\ref{prop:HK-eigenstates}
below): for a cube $\Gamma$, every normalized eigenstate
$\varphi\in\mathcal H_N$ of $H_0^{\rm fF}$ satisfies
\[
\left\langle
\varphi,(N_\Gamma-N\mu)^2\varphi
\right\rangle
\leq C_d NL^{d-1}(\log L)^{2d},
\]
where $C_d$ depends only on the dimension. Our bound removes
the logarithmic factor and holds for arbitrary subsets $\Gamma$.

In one dimension, \eqref{eq:intro-main-bound} removes the logarithmic
factor and the restriction to prime $L$ in~\cite{Roos2025}.
For intervals $\Gamma$, we also prove
\beq
\left|
\langle\varphi,N_\Gamma\varphi\rangle
-N\mu
\right|
\leq
C(1+\log N).
\label{sharp-mars}
\eeq

The strategy of our proof is as follows. We decompose
$N_\Gamma-N\mu$ with respect  to the one-particle energy transfer.
Applied to an energy eigenstate, components with distinct transfers
give orthogonal vectors. The contribution from nonzero transfers
is then controlled by the maximal number of momentum pairs
producing the same gap. With the notation of
Section~\ref{sec-free-fermion}, we prove the optimal bound
\begin{equation}
\label{eq:intro-resonance-bound}
\sup_{\Delta\neq0}
\#\left\{
(\mathbf k,\mathbf q)\in K_{L,d}^2:
\varepsilon(\mathbf k)-\varepsilon(\mathbf q)=\Delta
\right\}
\leq
C_dL^{d-1}.
\end{equation}
To prove this bound, we compare two representations of the same
gap, obtaining a vanishing sum of roots of unity. Mann's theorem~\cite{Mann1965} gives only finitely many
possibilities for each minimal vanishing subsum, up to rotation. A graph describes the constraints imposed by
inverse pairs of roots. Counting its connected components gives
\eqref{eq:intro-resonance-bound}.

Pairs of momenta with equal energies require a separate argument.
For each fixed nonzero momentum increment $\mathbf r$, an elementary
trigonometric count gives at most $2L^{d-1}$ solutions of
$\varepsilon(\mathbf q+\mathbf r)=\varepsilon(\mathbf q)$.
Together with Parseval's identity, this bounds their contribution
to the second moment by $C_dN\mu(1-\mu)L^{d-1}$ and completes the proof
of \eqref{eq:intro-variance-bound}.

The one-dimensional proof of Roos et al.~\cite{Roos2025} uses the
prime-size assumption to describe the many-body energy eigenspaces:
all degeneracies arise from changes of signs of the occupied momenta.
They then estimate the mean and variance of $N_\Gamma$ for arbitrary
superpositions within these eigenspaces. Our proof avoids this
description by working directly with one-particle energy transfers.

Hara and Koike~\cite{HaraKoike2026} reduce the time-averaged
density fluctuations to counting one-particle energy differences
in short intervals. They estimate these counts as a function of
the momentum transfer and combine the resulting bounds with
the Fourier coefficients of the box indicator to obtain the
ballistic equilibration scale. For an energy eigenstate, distinct energy transfers give
orthogonal vectors, leading to the short proof of
Theorem~\ref{prop:HK-eigenstates}.
To obtain  the sharp
bound, we instead control exact nonzero energy gaps over all
momentum pairs, removing the logarithmic loss and the restriction
on the geometry of $\Gamma$. We do not address finite-time equilibration in this paper.

After completing   this work, we became aware of the paper by
Xie, Zhao, and Zhao~\cite{Xie2024}, who study eigenvalue
multiplicities of the periodic discrete Laplacian using
vanishing sums of roots of unity.
 Their two-dimensional result implies the optimal
bound $24$ for nonzero one-particle gap multiplicities in dimension
one. In higher dimensions, their global multiplicity
bound~\cite[Corollary~5.8]{Xie2024} gives $O(L^d)$
when applied to our gap problem.
Their proof uses Evertse's theorem; we use Mann's theorem
and a graph-theoretic counting argument to obtain $O(L^{d-1})$
for nonzero gaps. We include a proof of the special case of
Mann's theorem needed here. Related results on sums of cosines
of rational angles appear in~\cite{Xue2026}.

The paper is organized as follows. Section~\ref{sec-free-fermion}
introduces the model and notation, and Section~\ref{sec-main}
states the main results. The resonance estimate is proved in
Section~\ref{sec-resonances}, and the main theorem in
Section~\ref{sec-op-red}. Section~\ref{sec-higher-optimality}
proves optimality in every dimension.
Section~\ref{sec-extensions} discusses extensions to Dirichlet
boundary conditions and to quadratic observables.
Section~\ref{sec-one-dimensional} proves the additional
one-dimensional expectation estimate. 
In Appendix~\ref{app:mann}, we prove the special case of Mann's
theorem used to establish \eqref{eq:intro-resonance-bound}.

\paragraph*{Acknowledgments.}
This work was partly funded by the CY Initiative grant
\emph{Investissements d'Avenir}, grant number ANR-16-IDEX-0008.
V.J. acknowledges the support of the MUR grant
``Dipartimento di Eccellenza 2023--2027'' of the Dipartimento di Matematica,
Politecnico di Milano.
We also acknowledge the support of the ANR project DYNACQ,
grant number ANR-24-CE40-5714.

{\emergencystretch=1em
C.V. gratefully acknowledges the financial support from the European Research
Council via the ERC Consolidator Grant RAMBAS (Project Nr. 101044249) and the
Deutsche Forschungsgemeinschaft (DFG, German Research Foundation) -- TRR 352,
Project-ID 470903074.\par}

The results presented here were developed during the Winter 2026 Mathematical
Physics Working Seminar at the Politecnico di Milano.
We thank A.~Lucia, M.~Moscolari, and M.~Wrochna for valuable discussions.
We also thank D.~Roy for advice on the number theory involved
in this work.

\section{The Periodic Free-Fermion Model}
\label{sec-free-fermion}
Throughout the paper, $C_d$ denotes a positive constant
depending only on $d$, whose value may change from line to line.

 Fix $d\geq1$ and let
\[
\Lambda=\Lambda_{L,d}=(\ZZ/L\ZZ)^d,
\qquad
V=|\Lambda|=L^d.
\]
The one-particle Hilbert space is $\mathfrak h=\ell^2(\Lambda)$, and
$I_{\mathfrak h}$ denotes its identity operator. Let
\[
\mathcal F
=
\bigoplus_{n=0}^{V}\wedge^n\mathfrak h
\]
be the fermionic Fock space over $\mathfrak h$, let $\Phi_{\rm vac}$ be the
Fock vacuum, and let $c_{\mathbf x},c_{\mathbf x}^*$,
$\mathbf x\in\Lambda$, be the fermionic annihilation and creation operators.
The particle-number operator is
\[
\mathcal N
=
\sum_{\mathbf x\in\Lambda}c_{\mathbf x}^*c_{\mathbf x}.
\]

The nearest-neighbour periodic free-fermion Hamiltonian is
\[
H_0^{\rm fF}
=
-\sum_{\mathbf x\in\Lambda}
\sum_{\nu=1}^d
\bigl(
 c_{\mathbf x}^*c_{\mathbf x+\mathbf e_\nu}
+c_{\mathbf x+\mathbf e_\nu}^*c_{\mathbf x}
\bigr),
\]
where $\mathbf e_\nu$ denotes the $\nu$th coordinate vector in $\mathbb Z^d$
and the coordinates are understood modulo $L$.
The Hamiltonian preserves
particle number, and throughout we restrict to the
$N$-particle sector
\[
\mathcal H_N
=
\ker(\mathcal N-N)
=
\wedge^N\mathfrak h,
\qquad
0<N<V.
\]

The dual lattice is
\[
K=K_{L,d}
=
\left\{
\frac{2\pi}{L}\mathbf m:
\mathbf m\in\{0,\ldots,L-1\}^d
\right\}.
\]
For $\mathbf k\in K$, define
\[
a_{\mathbf k}^*
=
\frac1{\sqrt V}
\sum_{\mathbf x\in\Lambda}
\e^{\i\mathbf k\cdot\mathbf x}c_{\mathbf x}^*,
\]
with inverse Fourier transform
\[
c_{\mathbf x}^*
=
\frac1{\sqrt V}
\sum_{\mathbf k\in K}
\e^{-\i\mathbf k\cdot\mathbf x}a_{\mathbf k}^*,
\qquad
c_{\mathbf x}
=
\frac1{\sqrt V}
\sum_{\mathbf k\in K}
\e^{\i\mathbf k\cdot\mathbf x}a_{\mathbf k}.
\]
The one-particle dispersion relation is
\[
\varepsilon(\mathbf k)
=
-2\sum_{\nu=1}^d\cos k_\nu,
\]
and hence
\[
H_0^{\rm fF}
=
\sum_{\mathbf k\in K}
\varepsilon(\mathbf k)a_{\mathbf k}^*a_{\mathbf k}.
\]
An $N$-particle configuration is an $N$-element subset
$X\subset K$. Fix an ordering of $K$, list the elements of
$X$ as $\mathbf k_1,\ldots,\mathbf k_N$ in this order, and set
\[
\Psi_X
=
a_{\mathbf k_1}^*\cdots a_{\mathbf k_N}^*\Phi_{\rm vac}.
\]
The vectors $\{\Psi_X\,|\,|X|=N\}$ form the standard Slater basis of
$\mathcal H_N$ and satisfy
\[
H_0^{\rm fF}\Psi_X
=
E(X)\Psi_X,
\qquad
E(X)
=
\sum_{\mathbf k\in X}\varepsilon(\mathbf k).
\]

Let $\Gamma\subset\Lambda$  and set
\[
N_\Gamma
=
\sum_{\mathbf x\in\Gamma}c_{\mathbf x}^*c_{\mathbf x},
\qquad
\mu=\frac{|\Gamma|}{V}.
\]
No regularity assumption on $\Gamma$ is needed for the main theorem. Recall that for 
$\eta>0$, $P_{{\rm neq},\eta}$ denotes the spectral projection of $N_\Gamma$
corresponding to $\{ |N_\Gamma/N-\mu|>\eta\}$. 

\section{Main Results}
\label{sec-main}
When $d=1$, Theorem~2 of \cite{Roos2025} takes the following form.

\begin{theorem}\label{thm:RSTTV}
Let $L$ be prime, let $46\leq N<L$, let $\Gamma\subset\Lambda$ be an
interval, and let $\eta>0$. Then every normalized eigenstate
$\varphi\in\mathcal H_N$ of $H_0^{\rm fF}$ satisfies
\[
\|P_{{\rm neq},\eta}\varphi\|^2
\leq
\frac{32\log N}{N\eta^2}.
\]
\end{theorem}

\begin{remark}
\label{rem:previous-constants}
The lower bound $N\geq46$ is a byproduct of the final numerical
simplification in the proof of \cite{Roos2025}. Their argument first
establishes an explicit variance estimate valid for all $N$, and the
restriction $N\geq46$ is introduced only to replace it by the simpler bound
$8N\log N$. The primality of $L$ is used in the explicit description of the
many-body eigenspaces. The condition $N<L/4$ appears in the original
statement of the theorem but is not used in its proof. The original
statement also assumes $\eta>2(\log N+1)/N$.
This restriction is inessential for the displayed estimate: below this
threshold its right-hand side is larger than one, while
$\|P_{{\rm neq},\eta}\varphi\|^2\leq1$.
$\diamond$
\end{remark}

Adapting Hara and Koike's proof of Proposition~2~\cite{HaraKoike2026}
gives the following eigenstate bound. We include a self-contained 
proof that makes the dependence on the particle number explicit.

\begin{theorem}
\label{prop:HK-eigenstates} Let $d\geq1$, $L\geq2$, $0<N<L^d$, and let
$\Gamma\subset\Lambda_{L,d}$ be a cube.
There exists a constant $C_d<\infty$, depending only on $d$,
such that every normalized eigenstate $\varphi\in\mathcal H_N$ of
$H_0^{\rm fF}$ satisfies
\[
\left\langle
\varphi,(N_\Gamma-N\mu)^2\varphi
\right\rangle
\leq C_dNL^{d-1}(\log L)^{2d}.
\]
Consequently, for every $\eta>0$,
\[
\|P_{{\rm neq},\eta}\varphi\|^2
\leq
\frac{C_d L^{d-1}(\log L)^{2d}}{N\eta^2}.
\]
\end{theorem}
\proof For $\mathbf r\in K$ set
\[
F_{\mathbf r}
=\sum_{\mathbf q\in K}a_{\mathbf q+\mathbf r}^*a_{\mathbf q},
\qquad
b_\Gamma(\mathbf r)
=\frac1V\sum_{\mathbf x\in\Gamma}\e^{-\i\mathbf r\cdot\mathbf x}.
\]
In the notation of Section~\ref{sec-op-red},
$b_\Gamma(\mathbf r)=V^{-1}\widehat\chi_\Gamma(-\mathbf r)$.
Since $b_\Gamma(0)=\mu$ and $F_0=N$ on $\mathcal H_N$, the momentum representation \eqref{eq:number-momentum} 
gives
\[
N_\Gamma-N\mu
=\sum_{\mathbf r\neq0}b_\Gamma(\mathbf r)F_{\mathbf r}.
\]
Fix $\mathbf r\neq0$ and set 
\[
S_{\mathbf r,\Delta}
=
\left\{\mathbf q\in K:
\varepsilon(\mathbf q+\mathbf r)-\varepsilon(\mathbf q)=\Delta
\right\}.
\]
Choose a coordinate $\nu$ for which
$r_\nu\not\equiv0\pmod{2\pi}$. Fix all coordinates of $\mathbf q$ except $q_\nu$. Since
\[
\cos(q_\nu+r_\nu)-\cos q_\nu
=-2\sin(r_\nu/2)\sin(q_\nu+r_\nu/2)
\]
and $\sin(r_\nu/2)\neq0$, the equation defining
$S_{\mathbf r,\Delta}$ reduces to
$\sin(q_\nu+r_\nu/2)=c$ for some constant $c$, 
and has at most two solutions modulo $2\pi$.
 Consequently, $|S_{\mathbf r,\Delta}|\leq 2L^{d-1}$. 

Write $H_0^{\rm fF}\varphi=E\varphi$. The commutator identity
\[
[H_0^{\rm fF},a_{\mathbf q+\mathbf r}^*a_{\mathbf q}]
=
\bigl(\varepsilon(\mathbf q+\mathbf r)-\varepsilon(\mathbf q)\bigr)
a_{\mathbf q+\mathbf r}^*a_{\mathbf q}
\]
shows that
$\sum_{\mathbf q\in S_{\mathbf r,\Delta}}
a_{\mathbf q+\mathbf r}^*a_{\mathbf q}\varphi$
belongs to the energy eigenspace with eigenvalue $E+\Delta$.
These vectors are  orthogonal for distinct $\Delta$.
By the Cauchy--Schwarz inequality,
\[
\|F_{\mathbf r}\varphi\|^2
=
\sum_\Delta
\left\|
\sum_{\mathbf q\in S_{\mathbf r,\Delta}}
a_{\mathbf q+\mathbf r}^*a_{\mathbf q}\varphi
\right\|^2
\leq
2L^{d-1}\sum_{\mathbf q\in K}
\|a_{\mathbf q+\mathbf r}^*a_{\mathbf q}\varphi\|^2
\leq
2L^{d-1}\sum_{\mathbf q\in K}
\|a_{\mathbf q}\varphi\|^2
=2NL^{d-1},
\]
where  the sum over $\Delta$ runs over the finitely many possible
energy transfers. The last inequality uses
$\|a_{\mathbf q+\mathbf r}^*\|\leq1$,
and the last equality uses
$\sum_{\mathbf q}a_{\mathbf q}^*a_{\mathbf q}=N$
on $\mathcal H_N$.
The triangle inequality now gives
\[
\left\langle\varphi,(N_\Gamma-N\mu)^2\varphi\right\rangle
\leq
2NL^{d-1}
\left(\sum_{\mathbf r\neq0}|b_\Gamma(\mathbf r)|\right)^2.
\]
For the reader's convenience, we give the elementary derivation
of the standard estimate for the Fourier coefficients of the cube
indicator. Let $\ell$ be the side length of the cube $\Gamma$. 
For $0\leq j<L$, set
\[
b_\ell(j)=\frac1L\sum_{x=0}^{\ell-1}\e^{-2\pi\i jx/L}.
\]
The absolute values $|b_\Gamma(\mathbf r)|$ are unchanged by
translations of $\Gamma$, and  we  may  assume that
$\Gamma=\{0,\ldots,\ell-1\}^d$. For
$\mathbf r=2\pi(j_1,\ldots,j_d)/L$,
\[
|b_\Gamma(\mathbf r)|
=\prod_{\nu=1}^d|b_\ell(j_\nu)|.
\]
We have $b_\ell(0)=\ell/L\leq1$, and for $1\leq j<L$,
summing the geometric series gives
\[
|b_\ell(j)|
=
\frac{|\sin(\pi\ell j/L)|}{L|\sin(\pi j/L)|}
\leq
\frac{1}{2\min\{j,L-j\}},
\]
where we used $\sin(\pi t)\geq2t$ for $0\leq t\leq1/2$.
Pairing the indices $j$ and $L-j$ gives
\[
\sum_{j=0}^{L-1}|b_\ell(j)|
\leq
1+\sum_{j=1}^{\lfloor L/2\rfloor}\frac1j
\leq C\log L
\]
for $L\geq 2$. Hence, 
\[
\sum_{\mathbf r\in K}|b_\Gamma(\mathbf r)|
=
\left(\sum_{j=0}^{L-1}|b_\ell(j)|\right)^d
\leq C^d(\log L)^d.
\]
This proves the first estimate. The second  one follows from Chebyshev's inequality.
\qed

\begin{remark}
\label{rem:HK-logarithmic-loss}
\emergencystretch=1em
The logarithmic factor in Theorem~\ref{prop:HK-eigenstates} comes from
the absolute summation of the Fourier coefficients of the cube
indicator, followed by squaring. In~\cite[Section~4.3]{HaraKoike2026},
the products $a_{\mathbf q+\mathbf r}^*a_{\mathbf q}$ are estimated
using the bound $\|a_{\mathbf q+\mathbf r}^*a_{\mathbf q}\|\leq1$.
The proof above instead retains
$\sum_{\mathbf q\in K}\|a_{\mathbf q}\varphi\|^2=N$, which gives the
explicit dependence on the particle number.\footnote{The second-moment
bounds of Theorems~\ref{prop:HK-eigenstates} and~\ref{thm:main} also
extend to normalized energy eigenstates $\varphi\in\mathcal F$ without
a fixed particle number. The nonzero components of $\varphi$ in the
particle-number sectors are eigenvectors of the same energy. Applying
the bounds to their normalizations and summing with the squared norms
as weights replaces $N_\Gamma-N\mu$ by $N_\Gamma-\mu\mathcal N$ and
$N$ on the right-hand side by
$\langle\varphi,\mathcal N\varphi\rangle$; the empty and fully occupied
sectors contribute zero. Since this expectation is at most $L^d$, the
resulting bounds are $C_d L^{2d-1}(\log L)^{2d}$ and
$C_d\mu(1-\mu)L^{2d-1}$, respectively, with the same assumptions on $\Gamma$.}
$\diamond$
\end{remark}
\begin{remark}
Hara and Koike~\cite{HaraKoike2026} establish equilibration on the
optimal ballistic scale $L$: at fixed positive particle density,
the fraction of nonequilibrium times tends to zero on windows
of duration $\mathcal T_L$ with $\mathcal T_L/L\to\infty$,
uniformly in the initial state. The eigenstate proof above does
not give this finite-time control. For general initial states,
distinct energy transfers need not give orthogonal vectors;
the  proof in~\cite{HaraKoike2026} handles the cross terms by time averaging and
counting one-particle energy differences in short intervals.
$\diamond$
\end{remark}



Our main result removes the logarithmic losses in Theorems~\ref{thm:RSTTV} and~\ref{prop:HK-eigenstates}, 
and holds for arbitrary subsets $\Gamma$.

\begin{theorem}\label{thm:main}
For every fixed $d\geq1$ there exists a constant $C_d<\infty$ such that,
for every side length $L$, every particle number $0<N<L^d$, and every
$\Gamma\subset\Lambda_{L,d}$, every normalized eigenstate
$\varphi\in\mathcal H_N$ of $H_0^{\rm fF}$ satisfies
\[
\left\langle
\varphi,(N_\Gamma-N\mu)^2\varphi
\right\rangle
\leq
C_dN\mu(1-\mu)L^{d-1}.
\]
Consequently, for every $\eta>0$,
\[
\|P_{{\rm neq},\eta}\varphi\|^2
\leq
\frac{C_d\mu(1-\mu)L^{d-1}}{N\eta^2}.
\]
\end{theorem}


The second-moment bound in Theorem~\ref{thm:main} is optimal
in every dimension. The concentration estimate is also optimal
as a bound uniform in $\eta$. Both statements are proved in
Section~\ref{sec-higher-optimality}.



In one dimension, we also obtain a sharper bound on the deviation
of the expected particle number from its equilibrium value.

\begin{theorem}\label{thm:one-dimensional-expectation}
Let $d=1$ and let $\Gamma\subset\Lambda$ be an interval. There exists a
universal constant $C<\infty$ such that every normalized eigenstate
$\varphi\in\mathcal H_N$ satisfies
\[
\left|
\langle\varphi,N_\Gamma\varphi\rangle
-N\mu
\right|
\leq
C(1+\log N).
\]
\end{theorem}

\section{One-Particle Resonances}
\label{sec-resonances}
For $\Delta\neq0$, define the one-particle gap multiplicity
\[
r_{L,d}(\Delta)
=
\#\left\{
(\mathbf k,\mathbf q)\in K_{L,d}^2:
\varepsilon(\mathbf k)-\varepsilon(\mathbf q)=\Delta
\right\}.
\]
The arithmetic input to the proof of Theorem~\ref{thm:main} is the following
sharp estimate.

\begin{theorem}\label{thm:res}
For every fixed $d\geq1$ there exists $R_d<\infty$ such that, for every
$L$ and every $\Delta\neq0$,
\beq
r_{L,d}(\Delta)
\leq
R_dL^{d-1}.
\label{key-key}
\eeq
Moreover, for every $L\geq2$ there exists a nonzero gap $\Delta$ for which $r_{L,d}(\Delta)\geq L^{d-1}$. 
\end{theorem}

The proof of the lower bound is elementary. Choose $p=0$ and
$q=2\pi/L$, so that $\cos p\neq\cos q$ for $L\geq2$. For each $\mathbf u\in K_{L,d-1}$, set
$\mathbf k=(p,\mathbf u)$ and $\mathbf q=(q,\mathbf u)$.
Here and below, $K_{L,0}$ is understood as a one-point set.
Then
\[
\varepsilon(\mathbf k)-\varepsilon(\mathbf q)
=-2\cos p+2\cos q\neq0,
\]
independently of $\mathbf u$. The $L^{d-1}$ choices of
$\mathbf u$  give distinct ordered pairs with the same
nonzero gap. 

The proof of the upper bound~\eqref{key-key} has two parts.
We first recall a theorem of Mann on vanishing sums of roots
of unity. We then apply it to the gap equation and use a
counting argument to bound the number of momentum pairs
representing the same nonzero gap.

\subsection{Minimal vanishing relations and Mann's theorem}

A relation $\zeta_1+\cdots+\zeta_m=0$ 
among roots of unity is called \emph{minimal} if no proper  subsum
vanishes. Its \emph{weight} is the number $m$ of terms. Two minimal relations are called rotationally equivalent if one is
obtained from the other by multiplying all roots by the same root of unity. 
The distinction between a relation and its rotational class is essential:
finitely many rotational classes may give rise to a number of
relations that grows with $L$.

We will use the following special case of Mann's
theorem~\cite{Mann1965}. For completeness, we give a proof
in Appendix~\ref{app:mann}. The \emph{order} of a root of
unity $\zeta$ is the smallest positive integer $n$ such
that $\zeta^n=1$.
We write $\mu_n$ for the multiplicative group of $n$th roots
of unity.

\begin{theorem}\label{thm:mann}
Let $\zeta_1+\cdots+\zeta_m=0$ 
be a minimal relation among roots of unity. After a common rotation, every
$\zeta_j$ has order dividing
\[
P_m=\prod_{p\leq m}p,
\]
where the product is over primes. In particular, for fixed $m$ there are
only finitely many minimal relations up to rotation.
\end{theorem}

\begin{remark}
Conway and Jones~\cite{ConwayJones1976} sharpened Mann's
restriction on the possible orders and classified minimal
relations of small weight. Poonen and
Rubinstein~\cite[Theorem~3]{PoonenRubinstein1998} extended
this classification to weight at most $12$.
W{\l}odarski's classification of vanishing sums of four
cosines~\cite{Wlodarski1969}, or Conway and Jones's classification
for eight roots of unity, suffices for the sharp multiplicity
bound of Xie, Zhao, and Zhao~\cite{Xie2024}, which gives
the bound $24$ for nonzero gaps in dimension one. Evertse~\cite{Evertse1999} bounds the number of nondegenerate
solutions in roots of unity of
$a_1z_1+\cdots+a_nz_n=1$, with fixed nonzero complex
coefficients, by a constant depending only on $n$.
$\diamond$
\end{remark}

\subsection{The resonance bound}
 We now use Mann's theorem to prove the upper bound in
Theorem~\ref{thm:res}. The strategy is the following. We fix one representation
of a nonzero gap and compare it with every other representation.
Subtracting the two gives a vanishing relation among roots of unity,
which we decompose into minimal vanishing blocks. Mann's theorem
leaves only finitely many possibilities for each block up to a
common rotation. The roots in the gap equation occur in pairs
whose product is $1$. These pairings impose relations between
the rotations of different blocks. We describe these relations
by a graph and use it to bound the number of momentum pairs
by $C_dL^{d-1}$.

\noindent{\bf Proof of the upper bound in Theorem~\ref{thm:res}.} Fix a nonzero gap
$\Delta=\varepsilon(\mathbf a)-\varepsilon(\mathbf b)$, with
$\mathbf a,\mathbf b\in K_{L,d}$. Let
$\zeta=\e^{2\pi\i/L}$. If
$\varepsilon(\mathbf k)-\varepsilon(\mathbf q)=\Delta$, then
\begin{equation}
\label{eq:resonance-root-relation}
\sum_{\nu=1}^d
\left(
\zeta^{k_\nu}+\zeta^{-k_\nu}
-\zeta^{q_\nu}-\zeta^{-q_\nu}
-\zeta^{a_\nu}-\zeta^{-a_\nu}
+\zeta^{b_\nu}+\zeta^{-b_\nu}
\right)
=0.
\end{equation}
Here and below, we write each momentum coordinate as
$2\pi j/L$ and use its integer label $j\in\ZZ/L\ZZ$
as the exponent of $\zeta$. Since the negative of a root of unity is again a root of unity,
we may absorb every minus sign into the corresponding root
and write \eqref{eq:resonance-root-relation} as a vanishing
relation of weight $8d$ among roots in $\mu_{2L}$.

\smallskip
\noindent
\emph{Minimal blocks and normalized shapes.} Label the $8d$ terms in \eqref{eq:resonance-root-relation}
by $1,\ldots,8d$ in their order of appearance, even when
some of the corresponding roots coincide.
Decompose the relation into minimal vanishing blocks by
successively removing minimal nonempty vanishing subsums.
This gives a partition of the label set $\{1,\ldots,8d\}$.
The number of possible partitions is bounded in terms of
$d$ alone.

Let $C$ be one of the blocks, let $m_C=|C|$, and denote its
labelled roots by $z_\ell$, $\ell\in C$. Choose the smallest
label $\ell_C\in C$ and set
\[
\rho_C=z_{\ell_C},
\qquad
\eta_\ell=\frac{z_\ell}{z_{\ell_C}},
\qquad \ell\in C.
\]
We call $(\eta_\ell)_{\ell\in C}$ the \emph{normalized shape}
of the block. It is obtained by a common rotation that makes
the root with label $\ell_C$ equal to $1$.
By Theorem~\ref{thm:mann}, each $\eta_\ell$ belongs to
$\mu_{P_{m_C}}$, and
\[
z_\ell=\rho_C\eta_\ell,
\qquad \ell\in C.
\]
Thus only finitely many normalized shapes are possible.
Since $m_C\leq8d$, their number is bounded in terms of $d$ alone.

We first count the momentum pairs that give a prescribed labelled
partition and prescribed normalized shapes for its blocks. After these finite choices have been made, the only
remaining quantities are the common rotations $\rho_C$.

\smallskip
\noindent
\emph{Anchored blocks.} We keep $\mathbf a$ and $\mathbf b$ fixed and count the pairs
$(\mathbf k,\mathbf q)$ representing the same gap.
For each coordinate $\nu$, we therefore call
$\zeta^{k_\nu}$, $\zeta^{-k_\nu}$, $-\zeta^{q_\nu}$, and
$-\zeta^{-q_\nu}$ the variable roots. The other four roots,
$-\zeta^{a_\nu}$, $-\zeta^{-a_\nu}$, $\zeta^{b_\nu}$, and
$\zeta^{-b_\nu}$, remain unchanged and will be called fixed.

We call a block \emph{anchored} if it contains at least one fixed
root. If $\ell\in C$ labels a fixed root, then $z_\ell$
is known and
\[
\rho_C=z_\ell\eta_\ell^{-1}.
\]
Thus the rotation of an anchored block is uniquely determined. If several
fixed roots belong to the same block, the remaining equalities
impose only compatibility conditions on the chosen normalized shape.

It remains to control the rotations of the unanchored blocks.

\smallskip
\noindent
\emph{The rotation graph.}
The variable roots come in inverse pairs, meaning that the
product of the two roots is $1$. These pairs impose relations
between the rotations of the blocks containing them, which
we describe by a graph.

Form a finite graph $G$, allowing multiple edges and loops,
whose vertices are the blocks containing at least one variable
root. A vertex is called anchored if its block is anchored.
For each $\nu=1,\ldots,d$, add one edge between the blocks
containing the labelled roots $\zeta^{k_\nu}$ and
$\zeta^{-k_\nu}$, and one edge between the blocks containing
$-\zeta^{q_\nu}$ and $-\zeta^{-q_\nu}$.
No edges are added for the fixed roots.
Since every block contains at least two of the $8d$ terms,
the number of vertices cannot exceed $4d$.
There are two edges for each $\nu$, so $G$ has exactly
$2d$ edges.
The degree of a vertex equals the number of variable terms
in its block: each term contributes one, and a loop
contributes two because both terms lie in the same block.

Consider an edge $e$ joining blocks $C$ and $D$, and let $\ell\in C$ and
$\ell'\in D$ label the two roots defining it. The
corresponding roots are $\rho_C\eta_\ell$ and
$\rho_D\eta_{\ell'}$, where $\eta_\ell$ and $\eta_{\ell'}$ are fixed by
the chosen normalized shapes. These roots form one of the inverse pairs
\[
\zeta^{k_\nu}\zeta^{-k_\nu}=1,
\qquad
(-\zeta^{q_\nu})(-\zeta^{-q_\nu})=1.
\]
Consequently, $(\rho_C\eta_\ell)(\rho_D\eta_{\ell'})=1$, 
and hence
\begin{equation}
\label{eq:rotation-edge-equation}
\rho_C\rho_D
=
(\eta_\ell\eta_{\ell'})^{-1}
=:
\gamma_e.
\end{equation}
The constant $\gamma_e$ is fixed by the chosen normalized
shapes and the labels $\ell,\ell'$.
If $e$ is a loop at $C$, the same argument gives
$\rho_C^2=\gamma_e$.
Once the rotation at one vertex has been chosen,
\eqref{eq:rotation-edge-equation} determines the rotations
at all other vertices in its connected component by successive
application of $\rho_D=\gamma_e/\rho_C$ along paths.
Additional edges impose compatibility conditions and cannot
increase the number of choices. Thus a component containing
an anchored vertex has at most one assignment of rotations.

For each unanchored component, fix a label $\ell$ belonging
to one of its blocks $C$. Since
$z_\ell=\rho_C\eta_\ell\in\mu_{2L}$, there are at most $2L$
choices for $z_\ell$. Each determines $\rho_C$ and hence all
rotations in the component. Every unanchored component therefore
has at most $2L$ assignments.

\smallskip
\noindent
\emph{Counting the components.} We now show that at most $d-1$ components can contribute a
factor $2L$ to the number of rotation assignments.

An unanchored component with exactly one edge has at most two
assignments. Indeed, if the edge joined two different vertices,
each of the two blocks would contain exactly one variable root.
Neither would contain a fixed root, so both would be one-term
vanishing blocks, which is impossible. The edge must therefore
be a loop, and its equation $\rho_C^2=\gamma_e$ has at most
two solutions.

To bound the number of unanchored components with at least two
edges, we use the following consequence of $\Delta\neq0$:
at least one block must contain both fixed and variable roots.
 Otherwise the blocks
containing variable roots would form a vanishing relation by
themselves, giving
\[
\sum_{\nu=1}^d
\left(
\zeta^{k_\nu}+\zeta^{-k_\nu}
-\zeta^{q_\nu}-\zeta^{-q_\nu}
\right)
=0.
\]
This would imply
$\varepsilon(\mathbf k)-\varepsilon(\mathbf q)=0$,
contradicting  $\Delta\neq0$. Thus $G$ has an anchored vertex with at least one incident edge.

Let $f$ be the number of unanchored connected components with
at least two edges. Together they contain at least $2f$ edges.
The graph $G$ has $2d$ edges in total, and at least one lies
outside these components because it belongs to a connected
component containing an anchored vertex. Thus $2f\leq2d-1$
and hence $f\leq d-1$.  Each of these $f$ components has at most $2L$ assignments. 
The anchored components have at most one assignment each,
and the unanchored one-edge components have at most two.
Since the number of components is bounded in terms of $d$ alone,
these contribute only a constant factor. Therefore,
for the fixed labelled partition and normalized shapes,
the total number of rotation assignments is bounded by
\[
C_d(2L)^f\leq C_dL^{d-1}.
\]

Since the normalized shapes are fixed, each assignment of rotations
determines every variable root through $z_\ell=\rho_C\eta_\ell$.
The roots $\zeta^{k_\nu}$ and $-\zeta^{q_\nu}$ then determine
$k_\nu$ and $q_\nu$ uniquely modulo $L$, and hence determine
$(\mathbf k,\mathbf q)$. Summing over the possible labelled
partitions and normalized block shapes, whose number is bounded
in terms of $d$ alone, gives  the upper bound. \qed

\section{Proof of the Main Theorem}
\label{sec-op-red}
\subsection{Matrix elements of the local particle-number operator}

Define
\[
\widehat\chi_\Gamma(\mathbf r)
=
\sum_{\mathbf x\in\Gamma}
\e^{\i\mathbf r\cdot\mathbf x}.
\]
The Fourier representation of $N_\Gamma$ is
\beq
N_\Gamma
=
\frac1V
\sum_{\mathbf k,\mathbf q\in K}
\widehat\chi_\Gamma(\mathbf q-\mathbf k)
 a_{\mathbf k}^*a_{\mathbf q}.
\label{eq:number-momentum}
\eeq
Set
\[
A_{\mathbf k\mathbf q}
=
\frac1V\widehat\chi_\Gamma(\mathbf q-\mathbf k),
\qquad
b=A-\mu I_{\mathfrak h}.
\]
Then $A$ is the one-particle orthogonal projection onto $\ell^2(\Gamma)$,
$A_{\mathbf k\mathbf k}=\mu$, and, on $\mathcal H_N$,
\[
B_\Gamma
:=
N_\Gamma-N\mu
=
\mathrm d\Gamma(b).
\]

\begin{lemma}\label{lem:matrix-elements}
Let $X,Y\subset K$ be $N$-particle configurations. Then
\[
\langle\Psi_X,N_\Gamma\Psi_Y\rangle=0
\]
unless $X=Y$, or $X$ and $Y$ differ in exactly one occupied momentum. More
precisely,
\[
\langle\Psi_X,N_\Gamma\Psi_X\rangle=N\mu,
\]
and if $X\setminus Y=\{\mathbf k\}$ and
$Y\setminus X=\{\mathbf q\}$, then
\[
\langle\Psi_X,N_\Gamma\Psi_Y\rangle
=
\pm A_{\mathbf k\mathbf q}.
\]
\end{lemma}

\proof
For $\mathbf k\neq\mathbf q$, the matrix element
\[
\langle\Psi_X,a_{\mathbf k}^*a_{\mathbf q}\Psi_Y\rangle
\]
can be nonzero only if $\mathbf q\in Y$, $\mathbf k\notin Y$, and
\[
X=(Y\setminus\{\mathbf q\})\cup\{\mathbf k\}.
\]
Thus an off-diagonal matrix element of $N_\Gamma$ can be nonzero only when
$X$ and $Y$ differ by a single replacement, in which case it equals
$\pm A_{\mathbf k\mathbf q}$. If $X=Y$, only the terms with
$\mathbf k=\mathbf q$ contribute, and each occupied momentum contributes
$A_{\mathbf k\mathbf k}=\mu$. This proves the claim.
\qed

\subsection{Energy-transfer decomposition}
Let $\Pi_E$ denote the spectral projection of $H_0^{\rm fF}$
corresponding to the eigenvalue $E$. For $\Delta\in\RR$, define
\[
B_\Delta
=
\sum_{\varepsilon(\mathbf k)-\varepsilon(\mathbf q)=\Delta}
 b_{\mathbf k\mathbf q}
 a_{\mathbf k}^*a_{\mathbf q}.
\]
Then
\[
B_\Gamma=\sum_\Delta B_\Delta,
\qquad
B_\Delta\operatorname{Ran}\Pi_E
\subset
\operatorname{Ran}\Pi_{E+\Delta}.
\]
Consequently, for every normalized
$\varphi\in\mathcal H_N\cap\operatorname{Ran}\Pi_E$,
\begin{equation}
\label{eq:transfer-orthogonality}
\|B_\Gamma\varphi\|^2
=
\sum_\Delta\|B_\Delta\varphi\|^2.
\end{equation}
We first estimate the nonzero transfers. By the Cauchy--Schwarz inequality
and Theorem~\ref{thm:res}, for any $\Delta\neq 0$ we have 
\[
\begin{split}
\|B_\Delta\varphi\|^2
&=
\left\|
\sum_{\varepsilon(\mathbf k)-\varepsilon(\mathbf q)=\Delta}
b_{\mathbf k\mathbf q}
a_{\mathbf k}^*a_{\mathbf q}\varphi
\right\|^2
\leq
\#\left\{
(\mathbf k,\mathbf q)\in K^2:
\varepsilon(\mathbf k)-\varepsilon(\mathbf q)=\Delta
\right\}
\sum_{\varepsilon(\mathbf k)-\varepsilon(\mathbf q)=\Delta}
|b_{\mathbf k\mathbf q}|^2
\|a_{\mathbf k}^*a_{\mathbf q}\varphi\|^2\\[2mm]
&\leq
R_dL^{d-1}
\sum_{\varepsilon(\mathbf k)-\varepsilon(\mathbf q)=\Delta}
|b_{\mathbf k\mathbf q}|^2
\|a_{\mathbf k}^*a_{\mathbf q}\varphi\|^2.
\end{split}
\]
Summing over $\Delta\neq0$, enlarging the sum over $\mathbf k$, and using
$\|a_{\mathbf k}^*\|=1$, we obtain
\[
\sum_{\Delta\neq0}\|B_\Delta\varphi\|^2
\leq
R_dL^{d-1}
\sum_{\mathbf q\in K}
\left(
\sum_{\mathbf k\in K}|b_{\mathbf k\mathbf q}|^2
\right)
\|a_{\mathbf q}\varphi\|^2.
\]
Since $A^2=A$ and $b=A-\mu I_{\mathfrak h}$,
\[
\sum_{\mathbf k\in K}|b_{\mathbf k\mathbf q}|^2
=
(b^2)_{\mathbf q\mathbf q}
=
\mu(1-\mu).
\]
Moreover, $\sum_{\mathbf q\in K}\|a_{\mathbf q}\varphi\|^2=N$, and so 
\begin{equation}
\label{eq:nonzero-transfer-bound}
\sum_{\Delta\neq0}\|B_\Delta\varphi\|^2
\leq
R_dN\mu(1-\mu)L^{d-1}.
\end{equation}
\subsection{The zero-transfer component}
Let $b_0$ be the one-particle matrix $(b_0)_{\mathbf k\mathbf q}=
\mathbf 1_{\{\varepsilon(\mathbf k)=\varepsilon(\mathbf q)\}}
 b_{\mathbf k\mathbf q}$.
Then $B_0=\mathrm d\Gamma(b_0)$. 

\begin{lemma}\label{lem:zero-transfer-hs}
For every $d\geq1$,\footnote{$\|b_0\|_{\rm HS}$ denotes the Hilbert-Schmidt norm of $b_0$.}
\[
\|b_0\|_{\rm HS}^2
\leq
2\mu(1-\mu)L^{d-1}.
\]
Consequently, on the $N$-particle sector,
\[
\|B_0\|^2
\leq
2N\mu(1-\mu)L^{d-1}.
\]
\end{lemma}

\proof
Since $b_{\mathbf q\mathbf q}=0$, only nonzero momentum increments occur.
For $\mathbf r\neq0$, set
\[
M(\mathbf r)
=
\#\left\{
\mathbf q\in K:
\varepsilon(\mathbf q+\mathbf r)=\varepsilon(\mathbf q)
\right\}.
\]
We repeat the counting argument from the proof of
Theorem~\ref{prop:HK-eigenstates}, here with $\Delta=0$.
Choose a coordinate $\nu$ such that
$r_\nu\not\equiv0\pmod{2\pi}$. The equality of energies
can be written as
\[
\sum_{j=1}^d
\sin\frac{r_j}{2}
\sin\left(q_j+\frac{r_j}{2}\right)
=0.
\]
After fixing the $d-1$ coordinates $q_j$, $j\neq\nu$, this is one sine
equation in $q_\nu$, and hence has at most two solutions modulo $2\pi$.
Thus $M(\mathbf r)\leq2L^{d-1}$. 
It follows that
\[
\|b_0\|_{\rm HS}^2
=
\sum_{\mathbf r\neq0}
M(\mathbf r)
\left|
\frac1V\widehat\chi_\Gamma(\mathbf r)
\right|^2
\leq
2L^{d-1}
\sum_{\mathbf r\neq0}
\left|
\frac1V\widehat\chi_\Gamma(\mathbf r)
\right|^2.
\]
Parseval's identity gives
\[
\sum_{\mathbf r\neq0}
\left|
\frac1V\widehat\chi_\Gamma(\mathbf r)
\right|^2
=
\mu(1-\mu),
\]
and proves the first claim.

Let $\lambda_1,\ldots,\lambda_V$ be the eigenvalues of the
self-adjoint operator $b_0$. The eigenvalues of $\mathrm d\Gamma(b_0)$ on
$\wedge^N\mathfrak h$ are the sums
$\lambda_{j_1}+\cdots+\lambda_{j_N}$ with
$1\leq j_1<\cdots<j_N\leq V$.
 Therefore
\[
\|\mathrm d\Gamma(b_0)\|
\leq
\sqrt N
\left(
\sum_{j=1}^V\lambda_j^2
\right)^{1/2}
=
\sqrt N\,\|b_0\|_{\rm HS}.
\]
This proves the second claim.
\qed

\noindent{\bf Proof of Theorem~\ref{thm:main}. }
Combining \eqref{eq:transfer-orthogonality},
\eqref{eq:nonzero-transfer-bound}, and
Lemma~\ref{lem:zero-transfer-hs}, we obtain
\[
\left\langle
\varphi,(N_\Gamma-N\mu)^2\varphi
\right\rangle
=
\|B_\Gamma\varphi\|^2
\leq
C_dN\mu(1-\mu)L^{d-1}
\]
for every normalized
$\varphi\in\mathcal H_N\cap\operatorname{Ran}\Pi_E$.
The concentration estimate follows from Chebyshev's inequality.
\qed


\section{Optimality}
\label{sec-higher-optimality}

We now prove the optimality statements announced in the introduction.
We construct exact energy eigenstates that give matching lower
bounds for both the variance and the concentration estimate in
every dimension.\footnote{A one-dimensional example with a binomial distribution
of the particle number on the even sites is discussed
in~\cite{BarghathiHerdmanDelMaestro2018}.}

The strategy of the construction is the following. We first
construct a density matrix that commutes with the Hamiltonian
and all spatial translations, and for which the distribution
of $N_\Gamma$ can be computed explicitly. We then decompose it into joint energy and translation eigenstates
and select one that gives the lower bound for the concentration
estimate. Translation symmetry ensures
that its expected particle number in $\Gamma$ is $N\mu$, so the
same eigenstate also gives the variance lower bound.

Let $L$ be even and set
\[
M=L^{d-1},\qquad R=L/2,\qquad N=RM=L^d/2,
\qquad
\Gamma=\{\mathbf x\in\Lambda_{L,d}:x_1\text{ is even}\}.
\]
Then $\mu=1/2$. Write each momentum as $(p,\mathbf u)$, with
$p\in K_{L,1}$ and $\mathbf u\in K_{L,d-1}$, and partition
the values of  $p$  into the pairs $\{p_r,q_r\}$, where
\[
p_r=\frac{2\pi r}{L},\qquad q_r=p_r+\pi,
\qquad 0\leq r<R.
\]
Choose independent random vectors $z_r=(z_{r,1},z_{r,2})$
uniformly distributed on the unit sphere of $\mathbb C^2$, and set
\[
f_{r,\mathbf u}
=
\bigl(
z_{r,1}a_{(p_r,\mathbf u)}^*
+z_{r,2}a_{(q_r,\mathbf u)}^*
\bigr)\Phi_{\rm vac},
\qquad 0\leq r<R,\quad \mathbf u\in K_{L,d-1}.
\]
These $RM=N$ one-particle vectors are orthonormal. List them in
a fixed order as $f_1,\ldots,f_N$, and consider their  normalized
Slater determinant 
\[
\Psi_z(\mathbf x_1,\ldots,\mathbf x_N)
=
\frac1{\sqrt{N!}}
\det\bigl[f_j(\mathbf x_i)\bigr]_{i,j=1}^N.
\]
Set $\rho_L=\mathbb E\,|\Psi_z\rangle\langle\Psi_z|$
on $\mathcal H_N$. 

We first show that the density matrix $\rho_L$ commutes with the
Hamiltonian and all spatial translations. After computing the
distribution of $N_\Gamma$ in $\rho_L$ and establishing a tail
lower bound, we will use these commutation properties to select
the required energy eigenstates.

Write 
$\varepsilon(p,\mathbf u)=-2\cos p+\varepsilon_\perp(\mathbf u)$.
Under time evolution, the coefficients $z_r$ are replaced by
\[
\bigl(\e^{2\i t\cos p_r}z_{r,1},
      \e^{2\i t\cos q_r}z_{r,2}\bigr).
\]
The additional factor $\e^{-\i t\varepsilon_\perp(\mathbf u)}$
multiplies the entire vector $f_{r,\mathbf u}$ and therefore
contributes only an overall phase to $\Psi_z$.  Similarly, a spatial translation multiplies the two components
of $z_r$ by phases depending on $p_r$ and $q_r$,  while the momenta $\mathbf u$ contribute only an overall phase
to $\Psi_z$. Multiplying the components of each $z_r$ by these phases preserves
its uniform distribution on the unit sphere and the independence
of the vectors $z_r$. The overall phase of $\Psi_z$ cancels in
$|\Psi_z\rangle\langle\Psi_z|$. Hence $\rho_L$ commutes with $H_0^{\rm fF}$ and all spatial translations.

We now compute the distribution of $N_\Gamma$ in $\rho_L$ by first
fixing $z$ and considering the state $\Psi_z$, and then averaging
over $z$. Let $Q_\Gamma$ denote the one-particle projection onto $\Gamma$,
which acts by multiplication by the characteristic function
\[
\one_\Gamma(\mathbf x)=\frac{1+(-1)^{x_1}}2.
\]
We have $N_\Gamma=\mathrm d\Gamma(Q_\Gamma)$. 
Multiplication by $(-1)^{x_1}$ shifts $p$ by $\pi$ and leaves
$\mathbf u$ unchanged. Since
$q_r=p_r+\pi$, it interchanges
$a_{(p_r,\mathbf u)}^*\Phi_{\rm vac}$ and
$a_{(q_r,\mathbf u)}^*\Phi_{\rm vac}$.
Thus $Q_\Gamma$ preserves their span and has matrix
\[
\frac12
\begin{pmatrix}
1&1\\
1&1
\end{pmatrix}
\]
in this basis.
Define
\[
g_{r,\mathbf u}^{\pm}
=
\frac{a_{(p_r,\mathbf u)}^*\pm a_{(q_r,\mathbf u)}^*}{\sqrt2}
\Phi_{\rm vac}.
\]
Then
\[
Q_\Gamma g_{r,\mathbf u}^{+}=g_{r,\mathbf u}^{+},
\qquad
Q_\Gamma g_{r,\mathbf u}^{-}=0,
\]
and
\[
f_{r,\mathbf u}
=
\frac{z_{r,1}+z_{r,2}}{\sqrt2}\,g_{r,\mathbf u}^{+}
+
\frac{z_{r,1}-z_{r,2}}{\sqrt2}\,g_{r,\mathbf u}^{-}.
\]
By multilinearity of the determinant, $\Psi_z$ can be expanded as a
linear combination of orthonormal Slater determinants obtained by
choosing either $g_{r,\mathbf u}^{+}$ or $g_{r,\mathbf u}^{-}$ for
each $(r,\mathbf u)$. Since $N_\Gamma=\mathrm d\Gamma(Q_\Gamma)$, each determinant
is an eigenvector of $N_\Gamma$, with eigenvalue equal to
the number of $+$ choices. Its coefficient
is the product of the corresponding coefficients in the expansions
of the vectors $f_{r,\mathbf u}$. Since these determinants are
orthonormal, the probability of each eigenvalue is the sum of the
squared moduli of the coefficients of the determinants with that
eigenvalue.

For fixed $z$, this distribution can be described by independent
Bernoulli variables $X_{r,\mathbf u}$, where $X_{r,\mathbf u}=1$
corresponds to the $+$ choice and $X_{r,\mathbf u}=0$ to the $-$
choice, with
\[
\mathbb P(X_{r,\mathbf u}=1\mid z)
=T_r:=\frac{|z_{r,1}+z_{r,2}|^2}{2},
\qquad
\mathbb P(X_{r,\mathbf u}=0\mid z)=1-T_r.
\]
Set
\[
U_r=\sum_{\mathbf u\in K_{L,d-1}}X_{r,\mathbf u}.
\]
Conditional on $z$, the variables $U_r$ are independent and
\[
\mathbb P(U_r=j\mid z)
=\binom Mj T_r^j(1-T_r)^{M-j},
\qquad j=0,\ldots,M.
\]
The distribution of $N_\Gamma$ in $\Psi_z$ is therefore the
conditional distribution of $\sum_{r=0}^{R-1}U_r$.

To average this conditional distribution, we determine the law
of $T_r$. By unitary invariance,
\[
T_r=\left|\frac{z_{r,1}+z_{r,2}}{\sqrt2}\right|^2
\]
has the same law as $|z_{r,1}|^2$. Parametrize the unit sphere
of $\mathbb C^2$ by
\[
z_1=\cos\theta\,\e^{\i\alpha},
\qquad
z_2=\sin\theta\,\e^{\i\beta},
\qquad
0\leq\theta\leq\frac\pi2,\quad 0\leq\alpha,\beta<2\pi.
\]
The normalized surface measure is
\[
2\sin\theta\cos\theta\,\mathrm d\theta\,
\frac{\mathrm d\alpha}{2\pi}\,
\frac{\mathrm d\beta}{2\pi}.
\]
The change of variable $t=\cos^2\theta$  shows that
$|z_1|^2$, and hence $T_r$, is uniform on $[0,1]$.
 The variables $T_r$ are
independent because the $z_r$ are independent. Averaging the conditional binomial 
distribution over $T_r$ gives\footnote{The averaging used here can also be understood
through the standard identity for averages of tensor powers
of pure states; see~\cite[Proposition~6]{Harrow2013}.}
\[
\mathbb P(U_r=j)
=\binom Mj\int_0^1t^j(1-t)^{M-j}\,\mathrm dt
=\binom Mj\frac{j!(M-j)!}{(M+1)!}
=\frac1{M+1}.
\]
The variables $U_r$ are also independent after averaging. Indeed,
conditional independence and the independence of the $T_r$ give,
for $j_0,\ldots,j_{R-1}\in\{0,\ldots,M\}$,
\[
\mathbb P(U_0=j_0,\ldots,U_{R-1}=j_{R-1})
=\mathbb E\prod_{r=0}^{R-1}
\binom M{j_r}T_r^{j_r}(1-T_r)^{M-j_r}
=\prod_{r=0}^{R-1}\mathbb P(U_r=j_r).
\]
Consequently, the spectral distribution of $N_\Gamma$ in $\rho_L$
is that of $\sum_{r=0}^{R-1}U_r$, where the $U_r$ are independent
and uniformly distributed on $\{0,\ldots,M\}$.

The mean of $U_r$ is $M/2$. To compute its variance,  we condition on $T_r$:
\[
\begin{split}
\operatorname{Var}(U_r)
&=\mathbb E\bigl[\operatorname{Var}(U_r\mid T_r)\bigr]
  +\operatorname{Var}\bigl(\mathbb E[U_r\mid T_r]\bigr)\\
&=M\mathbb E\bigl[T_r(1-T_r)\bigr]
  +M^2\operatorname{Var}(T_r)\\
&=\frac M6+\frac{M^2}{12}
=\frac{M(M+2)}{12}.
\end{split}
\]
Conditional on $T_r$, the $M$ Bernoulli variables whose sum is $U_r$
are independent.  The term $M^2/12$ is the variance of the conditional mean
$\mathbb E[U_r\mid T_r]=MT_r$. When $M>1$, averaging over $T_r$ creates
correlations between these variables. The variables  $U_r$
for different $r$  remain independent, so their variances add.
We obtain
\begin{equation}
\label{eq:optimality-uniform-variance}
\operatorname{tr}(\rho_LN_\Gamma)=\frac N2,
\qquad
\sigma_L^2
:=\operatorname{tr}\!\left(\rho_L(N_\Gamma-N/2)^2\right)
=\frac{RM(M+2)}{12}
=\frac{N(M+2)}{12}.
\end{equation}
The term $RM^2/12$ has precisely the scale $NL^{d-1}$. 

To obtain a tail lower bound, we apply the Lindeberg--Feller
central limit theorem for triangular arrays
\cite[Section~XV.6]{Feller1971} to the independent variables
$U_r$, whose distributions may depend on $L$.
For each $r$,
\[
\frac{|U_r-M/2|}{\sigma_L}
\leq\frac{M}{2\sigma_L}
=\sqrt{\frac{3M}{R(M+2)}}
\leq\sqrt{\frac3R}\rightarrow0.
\]
The Lindeberg condition is therefore satisfied, and
\[
\frac{\sum_{r=0}^{R-1}U_r-N/2}{\sigma_L}
\xrightarrow[L\to\infty]{\rm d} Z,
\]
where $Z$ is a standard normal random variable.
Fix $s>0$ and set\footnote{Since $\eta_L\asymp L^{-1/2}\to0$, the sharpness
statement concerns concentration estimates uniform in $\eta$.}
\[
\eta_L=\frac{s\sigma_L}{N},
\qquad p_s=\mathbb P(|Z|>s)>0.
\]
Since the normal distribution has no atoms, $\operatorname{tr}(\rho_LP_{{\rm neq},\eta_L})
\rightarrow p_s$. 

We now select energy eigenstates from a decomposition of $\rho_L$.
The Hamiltonian, $\rho_L$, and the translation unitaries are commuting
normal operators on $\mathcal H_N$. Choose a common orthonormal
eigenbasis $\{\psi_j\}$ and write
\[
\rho_L=\sum_jw_j|\psi_j\rangle\langle\psi_j|,
\qquad w_j\geq0,\qquad\sum_jw_j=1.
\]
Since each $\psi_j$ is a translation eigenvector,
$\langle\psi_j,c_{\mathbf x}^*c_{\mathbf x}\psi_j\rangle$
is independent of $\mathbf x$.  Since the total particle number
is $N$, this gives
\[
\langle\psi_j,c_{\mathbf x}^*c_{\mathbf x}\psi_j\rangle
=\frac{N}{L^d},
\qquad
\langle\psi_j,N_\Gamma\psi_j\rangle=\frac N2.
\]
Hence $\langle\psi_j,(N_\Gamma-N/2)^2\psi_j\rangle
=\operatorname{Var}_{\psi_j}(N_\Gamma)$. Moreover,
\[
\operatorname{tr}(\rho_LP_{{\rm neq},\eta_L})
=\sum_jw_j\|P_{{\rm neq},\eta_L}\psi_j\|^2.
\]
For each $L$, at least one basis vector $\varphi_L$ satisfies
\[
\|P_{{\rm neq},\eta_L}\varphi_L\|^2
\geq\operatorname{tr}(\rho_LP_{{\rm neq},\eta_L}).
\]
The energy eigenstates $\varphi_L$ therefore satisfy
\begin{equation}
\label{eq:optimality-summary-tail}
\liminf_{L\to\infty}
\|P_{{\rm neq},\eta_L}\varphi_L\|^2\geq p_s>0.
\end{equation}
By the definition of $P_{{\rm neq},\eta_L}$ and the identity
$N\eta_L=s\sigma_L$,
\[
\operatorname{Var}_{\varphi_L}(N_\Gamma)
\geq s^2\sigma_L^2
\|P_{{\rm neq},\eta_L}\varphi_L\|^2.
\]
Since $\sigma_L^2/(NL^{d-1})=(M+2)/(12M)\geq1/12$, we obtain
\begin{equation}
\label{eq:optimality-summary-variance}
\liminf_{L\to\infty}
\frac{\operatorname{Var}_{\varphi_L}(N_\Gamma)}{NL^{d-1}}
\geq\frac{s^2p_s}{12}>0.
\end{equation}
Finally,
\[
\frac{N\eta_L^2}{L^{d-1}}
=\frac{s^2(M+2)}{12M}
\]
is bounded above and below by positive constants. Thus
$\eta_L\asymp\sqrt{L^{d-1}/N}$. We conclude that the factor $NL^{d-1}$
in the variance bound is optimal, as is the coefficient
$L^{d-1}/N$ in concentration estimates uniform in $\eta$.

For $L\geq4$, the region $\Gamma$ in this example is disconnected.
The same lower bound on the variance can also be obtained for
rectangular regions in every dimension. We omit these alternative
constructions for reasons of space.


\section{Extensions}
\label{sec-extensions}
The proof of Theorem~\ref{thm:main} uses the orthogonality of
$B_\Delta\varphi$ for distinct $\Delta$, the resonance bound
for nonzero transfers, and a Hilbert--Schmidt estimate for
the zero-transfer component. These arguments also give
the following two extensions.

\subsection{Dirichlet boundary conditions}
\label{subsec:dirichlet-extension}
Let $\Lambda^{\rm D}_{L,d}=\{1,\ldots,L\}^d$,
$\mathfrak h^{\rm D}=\ell^2(\Lambda^{\rm D}_{L,d})$, 
and let
\[
\mathcal F^{\rm D}
=
\bigoplus_{n=0}^{L^d}\wedge^n\mathfrak h^{\rm D}
\]
be the corresponding fermionic Fock space. Let $H_0^{{\rm D},{\rm fF}}$ be the nearest-neighbour free-fermion
Hamiltonian with Dirichlet boundary conditions. Its one-particle energies are
\[
\varepsilon^{\rm D}(\mathbf j)
=
-2\sum_{\nu=1}^d
\cos\frac{\pi j_\nu}{L+1},
\qquad
\mathbf j\in\{1,\ldots,L\}^d.
\]

For $0<N<L^d$, let $\mathcal H_N^{\rm D}=
\wedge^N\mathfrak h^{\rm D}$. 
Given $\Gamma\subset\Lambda^{\rm D}_{L,d}$, set $\mu=|\Gamma|/L^d$, $P_{{\rm neq},\eta}^{\rm D}
=
\mathbf 1_{\{|N_\Gamma/N-\mu|>\eta\}}
$. 

\begin{theorem}\label{thm:dirichlet}
For every fixed $d\geq1$ there exists $C_d<\infty$ such that, for every
side length $L$, every particle number $0<N<L^d$, every
$\Gamma\subset\Lambda^{\rm D}_{L,d}$, every $\eta>0$, and every normalized
eigenstate $\varphi\in\mathcal H_N^{\rm D}$ of
$H_0^{{\rm D},{\rm fF}}$,
\[
\left\langle
\varphi,(N_\Gamma-N\mu)^2\varphi
\right\rangle
\leq
C_dNL^{d-1}
\]
and
\[
\|P_{{\rm neq},\eta}^{\rm D}\varphi\|^2
\leq
\frac{C_dL^{d-1}}{N\eta^2}.
\]
\end{theorem}
\proof
Set $M=2(L+1)$ 
and let $\widetilde\Lambda=(\ZZ/M\ZZ)^d$. 
The Dirichlet one-particle energies satisfy
\[
\varepsilon^{\rm D}(\mathbf j)
=
-2\sum_{\nu=1}^d
\cos\left(\frac{2\pi j_\nu}{M}\right),
\qquad
\mathbf j\in\{1,\ldots,L\}^d.
\]
Thus the Dirichlet spectrum is contained in the spectrum of the periodic
one-particle Hamiltonian on $\widetilde\Lambda$. It follows from
Theorem~\ref{thm:res} that, for every $\Delta\neq0$,
\[
\#\left\{
(\mathbf j,\mathbf k)\in\{1,\ldots,L\}^{2d}:
\varepsilon^{\rm D}(\mathbf j)
-
\varepsilon^{\rm D}(\mathbf k)
=
\Delta
\right\}
\leq
C_dM^{d-1}
\leq
C_dL^{d-1}.
\]

Let $A^{\rm D}$ be the one-particle projection onto
$\ell^2(\Gamma)$ and set $b^{\rm D}=A^{\rm D}-\mu I_{\mathfrak h^{\rm D}}$. 
Decompose
\[
\mathrm d\Gamma(b^{\rm D})
=
\sum_\Delta B_\Delta
\]
according to the one-particle energy transfer. For a normalized eigenvector $\varphi\in\mathcal H_N^{\rm D}$,
the vectors $B_\Delta\varphi$ are pairwise orthogonal by the same
argument as in \eqref{eq:transfer-orthogonality}. Since $\|b^{\rm D}\|\leq1$, the argument proving
\eqref{eq:nonzero-transfer-bound}, with the Dirichlet energies in place
of the periodic ones, gives
\begin{equation}
\label{eq:dirichlet-nonzero-transfer}
\sum_{\Delta\neq0}
\|B_\Delta\varphi\|^2
\leq
C_dNL^{d-1}.
\end{equation}
It remains to estimate the zero-transfer component. Let
$\widetilde h_0$ be the periodic one-particle Hamiltonian on
$\ell^2(\widetilde\Lambda)$, and let $\mathfrak h_{\rm odd}$ be the
subspace consisting of functions which are odd under each coordinate
reflection $x_\nu\mapsto-x_\nu$ modulo $M$. Let $h_0^{\rm D}$ denote the one-particle Dirichlet Hamiltonian.
The normalized odd extension defines a unitary
$U:\mathfrak h^{\rm D}\longrightarrow\mathfrak h_{\rm odd}$
satisfying
\[
\widetilde h_0U=Uh_0^{\rm D}.
\]
Let $\widetilde\Gamma$ be the union of the $2^d$ coordinate reflections
of $\Gamma$. These reflected copies are disjoint, and hence
\[
|\widetilde\Gamma|
=
2^d|\Gamma|,
\qquad
\widetilde\mu
:=
\frac{|\widetilde\Gamma|}{M^d}
=
\frac{|\Gamma|}{(L+1)^d}.
\]
Moreover, $A^{\rm D}
=
U^*\one_{\widetilde\Gamma}U$. 
Set $\widetilde b
=
\one_{\widetilde\Gamma}
-
\widetilde\mu I_{\ell^2(\widetilde\Lambda)},
$
and $c
=
U^*\widetilde b\,U
=
A^{\rm D}
-
\widetilde\mu I_{\mathfrak h^{\rm D}}.
$

Let $Q$ denote the orthogonal projection onto $\mathfrak h_{\rm odd}$.
The set $\widetilde\Gamma$ is invariant under each coordinate reflection.
Consequently, both $\widetilde h_0$ and $\widetilde b$ commute with $Q$.
If $\widetilde b_0$ denotes the zero-transfer component of
$\widetilde b$ with respect to $\widetilde h_0$, and $c_0$ denotes the
zero-transfer component of $c$ with respect to the Dirichlet Hamiltonian,
then $c_0=U^*Q\widetilde b_0QU$. 
Therefore, $\|c_0\|_{\rm HS}^2
\leq
\|\widetilde b_0\|_{\rm HS}^2.
$
Applying Lemma~\ref{lem:zero-transfer-hs} on the torus of side $M$ gives
\[
\|c_0\|_{\rm HS}^2
\leq
C_dM^{d-1}
\leq
C_dL^{d-1}.
\]
Since $b^{\rm D}=c+(\widetilde\mu-\mu)I_{\mathfrak h^{\rm D}}$, we have
\[
(b^{\rm D})_0
=
c_0+
(\widetilde\mu-\mu)I_{\mathfrak h^{\rm D}}.
\]
Furthermore,
\[
|\widetilde\mu-\mu|
=
|\Gamma|
\left(
\frac1{L^d}
-
\frac1{(L+1)^d}
\right)
=
\mu
\left[
1-
\left(\frac{L}{L+1}\right)^d
\right]
\leq
\frac dL.
\]
It follows that
\[
\begin{split}
\|(b^{\rm D})_0\|_{\rm HS}^2
&\leq
2\|c_0\|_{\rm HS}^2
+
2|\widetilde\mu-\mu|^2
\dim\mathfrak h^{\rm D}\\
&\leq
C_dL^{d-1}
+
2d^2L^{d-2}\\
&\leq
C_dL^{d-1}.
\end{split}
\]
Since $B_0=\mathrm d\Gamma((b^{\rm D})_0)$, 
the second-quantization estimate used in
Lemma~\ref{lem:zero-transfer-hs} yields
\begin{equation}
\label{eq:dirichlet-zero-transfer}
\|B_0\|^2
\leq
N\|(b^{\rm D})_0\|_{\rm HS}^2
\leq
C_dNL^{d-1}.
\end{equation}
On $\mathcal H_N^{\rm D}$, $\mathrm d\Gamma(b^{\rm D})=
N_\Gamma-N\mu$. 
Using the orthogonality of $B_\Delta\varphi$ for distinct $\Delta$,
together with
\eqref{eq:dirichlet-nonzero-transfer} and
\eqref{eq:dirichlet-zero-transfer}, we conclude that
\[
\left\langle
\varphi,(N_\Gamma-N\mu)^2\varphi
\right\rangle
=
\left\|
\mathrm d\Gamma(b^{\rm D})\varphi
\right\|^2
=
\sum_\Delta\|B_\Delta\varphi\|^2
\leq
C_dNL^{d-1}.
\]
The concentration estimate follows from Chebyshev's inequality. \qed
\subsection{Quadratic observables}
\label{subsec:quadratic-extension}
We return to periodic boundary conditions. Let $h_0$ be the
one-particle Hamiltonian on $\mathfrak h=\ell^2(\Lambda_{L,d})$,
so that
\[
(h_0)_{\mathbf k\mathbf q}
=
\varepsilon(\mathbf k)\delta_{\mathbf k\mathbf q},
\qquad
H_0^{\rm fF}=\mathrm d\Gamma(h_0).
\]
For a self-adjoint operator $s=s_{L,d}$ on $\mathfrak h$, set
$\mathcal A=\mathrm d\Gamma(s)$.
Suppose that there exist real numbers $\alpha=\alpha_{L,d}$
and $\beta=\beta_{L,d}$ such that
\[
g=s-\alpha h_0-\beta I_{\mathfrak h}
\]
satisfies
\begin{equation}
\label{eq:quadratic-column-condition}
\sup_{\mathbf q\in K}
\sum_{\mathbf k\in K}|g_{\mathbf k\mathbf q}|^2
\leq C_g
\end{equation}
and
\begin{equation}
\label{eq:quadratic-zero-condition}
\|g_0\|_{\rm HS}^2
\leq C_gL^{d-1},
\qquad
(g_0)_{\mathbf k\mathbf q}
=
\one_{\{\varepsilon(\mathbf k)=\varepsilon(\mathbf q)\}}
g_{\mathbf k\mathbf q},
\end{equation}
where $C_g$ may depend on $d$ but is independent of $L$.

\begin{proposition}\label{prop:quadratic-general}
Under \eqref{eq:quadratic-column-condition} and
\eqref{eq:quadratic-zero-condition}, there exists a constant
$C_{d,g}$, depending only on $d$ and $C_g$, such that, for
every $0<N<L^d$ and every energy $E$, every normalized
$\varphi\in\mathcal H_N\cap\operatorname{Ran}\Pi_E$ satisfies
\[
\left\langle
\varphi,
(\mathcal A-\alpha E-\beta N)^2\varphi
\right\rangle
\leq
C_{d,g}NL^{d-1}.
\]
Consequently, for every $\eta>0$,
\[
\left\|
\mathbf 1_{\{|\mathcal A-\alpha E-\beta N|>N\eta\}}
\varphi
\right\|^2
\leq
\frac{C_{d,g}L^{d-1}}{N\eta^2}.
\]
\end{proposition}

\proof
On $\mathcal H_N\cap\operatorname{Ran}\Pi_E$,
\[
\mathcal A-\alpha E-\beta N
=
\mathrm d\Gamma(g).
\]
Let $G_\Delta$ denote the component of $\mathrm d\Gamma(g)$ with
one-particle energy transfer $\Delta$. Theorem~\ref{thm:res}, together
with \eqref{eq:quadratic-column-condition}, gives
\[
\sum_{\Delta\neq0}\|G_\Delta\varphi\|^2
\leq
R_dC_gNL^{d-1}.
\]
On $\mathcal H_N$, the zero-transfer component
$G_0=\mathrm d\Gamma(g_0)$ satisfies
\[
\|G_0\|^2
\leq
N\|g_0\|_{\rm HS}^2
\leq
C_gNL^{d-1}.
\]
The orthogonality of $G_\Delta\varphi$ for distinct $\Delta$
proves the second-moment estimate;
the concentration estimate follows from Chebyshev's inequality.
\qed

The particle-number operator $N_\Gamma$ is a special case of
Proposition~\ref{prop:quadratic-general}, with $s=A$,
$\alpha=0$, and $\beta=\mu$, where $A$ is the one-particle
projection onto $\ell^2(\Gamma)$. The two hypotheses were
verified in the proof of Theorem~\ref{thm:main}.

The proposition also applies to sums of translates of a fixed
local one-particle operator. Let $R\subset\ZZ^d$ be finite and
let $b_{\rm loc}$ be a fixed self-adjoint operator on $\ell^2(R)$.
For all sufficiently large $L$, identify $R$ with its image in
$\Lambda_{L,d}$ and extend $b_{\rm loc}$ by zero to $\mathfrak h$.
Let $\theta^{\mathbf x}$ denote the translation on $\mathfrak h$
defined by $\theta^{\mathbf x}\delta_{\mathbf y}
=\delta_{\mathbf y+\mathbf x}$, with coordinates taken modulo $L$.
For $\Gamma\subset\Lambda_{L,d}$, set
\[
s_\Gamma
=
\sum_{\mathbf x\in\Gamma}
\theta^{\mathbf x}b_{\rm loc}\theta^{-\mathbf x},
\qquad
\mathcal A_\Gamma=\mathrm d\Gamma(s_\Gamma),
\qquad
\mu_\Gamma=\frac{|\Gamma|}{V}.
\]

\begin{proposition}\label{prop:quadratic-local-translates}
There exists a constant $C_{d,b}<\infty$, depending only on $d$
and $b_{\rm loc}$, such that, for every sufficiently large $L$,
every $0<N<L^d$, every $\Gamma\subset\Lambda_{L,d}$, every energy
$E$, and every normalized
$\varphi\in\mathcal H_N\cap\operatorname{Ran}\Pi_E$,
\[
\left\langle
\varphi,
\left(\mathcal A_\Gamma
-\mu_\Gamma\mathcal A_{\Lambda_{L,d}}\right)^2
\varphi
\right\rangle
\leq C_{d,b}N\mu_\Gamma(1-\mu_\Gamma)L^{d-1}.
\]
\end{proposition}

\proof
Set $g=s_\Gamma-\mu_\Gamma s_{\Lambda_{L,d}}$.
In the momentum basis,
\[
(b_{\rm loc})_{\mathbf k\mathbf q}
=
\frac1V\sum_{\mathbf x,\mathbf y\in R}
\e^{\i(\mathbf q\cdot\mathbf y-\mathbf k\cdot\mathbf x)}
\langle\delta_{\mathbf x},b_{\rm loc}\delta_{\mathbf y}\rangle.
\]
Thus $|(b_{\rm loc})_{\mathbf k\mathbf q}|\leq C_b/V$, where
\[
C_b
=
\sum_{\mathbf x,\mathbf y\in R}
\left|\langle\delta_{\mathbf x},b_{\rm loc}\delta_{\mathbf y}\rangle\right|
\]
is independent of $L$. Translation gives
\[
g_{\mathbf k\mathbf q}
=
(b_{\rm loc})_{\mathbf k\mathbf q}
\left(
\widehat\chi_\Gamma(\mathbf q-\mathbf k)
-|\Gamma|\delta_{\mathbf k\mathbf q}
\right).
\]
In particular, $g_{\mathbf q\mathbf q}=0$. By Parseval's identity,
\[
\sup_{\mathbf q\in K}
\sum_{\mathbf k\in K}|g_{\mathbf k\mathbf q}|^2
\leq
\frac{C_b^2}{V^2}
\sum_{\mathbf r\neq0}|\widehat\chi_\Gamma(\mathbf r)|^2
=
C_b^2\mu_\Gamma(1-\mu_\Gamma).
\]
For each $\mathbf r\neq0$, the counting argument in
Lemma~\ref{lem:zero-transfer-hs} bounds the number of solutions
of $\varepsilon(\mathbf q+\mathbf r)=\varepsilon(\mathbf q)$
by $2L^{d-1}$. Hence
\[
\|g_0\|_{\rm HS}^2
\leq
\frac{2C_b^2L^{d-1}}{V^2}
\sum_{\mathbf r\neq0}|\widehat\chi_\Gamma(\mathbf r)|^2
=
2C_b^2\mu_\Gamma(1-\mu_\Gamma)L^{d-1}.
\]
For $0<\mu_\Gamma<1$, apply
Proposition~\ref{prop:quadratic-general} to
$s=g/\sqrt{\mu_\Gamma(1-\mu_\Gamma)}$ with
$\alpha=\beta=0$ and $C_g=2C_b^2$.
Multiplying the resulting bound by $\mu_\Gamma(1-\mu_\Gamma)$
gives the claim. If $\mu_\Gamma=0$ or $\mu_\Gamma=1$, then
$g=0$ and the claim is immediate.
\qed

If $\varphi$ is also an eigenvector of
$\mathcal A_{\Lambda_{L,d}}$ with eigenvalue $a$, the same
estimate holds with $\mathcal A_\Gamma-\mu_\Gamma a$ in place
of $\mathcal A_\Gamma-\mu_\Gamma\mathcal A_{\Lambda_{L,d}}$.

For $b_{\rm loc}=|\delta_{\mathbf 0}\rangle
\langle\delta_{\mathbf 0}|$, we have $\mathcal A_\Gamma=N_\Gamma$
and $\mathcal A_{\Lambda_{L,d}}=\mathcal N$. The choice
\[
b_{\rm loc}
=
-\sum_{\nu=1}^d
\left(
|\delta_{\mathbf 0}\rangle\langle\delta_{\mathbf e_\nu}|
+
|\delta_{\mathbf e_\nu}\rangle\langle\delta_{\mathbf 0}|
\right)
\]
gives $\mathcal A_{\Lambda_{L,d}}=H_0^{\rm fF}$.
In this case, $\mathcal A_\Gamma$ includes hopping terms across
the boundary of $\Gamma$.

We now apply Proposition~\ref{prop:quadratic-general} to the
energy contained in a spatial region, defined by keeping only
the hopping terms whose two endpoints belong to that region.

Let ${\cal I}\subset\ZZ/L\ZZ$ be an interval of length $\ell$,
where $1\leq\ell<L$, let $\Gamma={\cal I}^d$, and set
\[
\mathcal B_{\Gamma,\nu}
=
\{\mathbf x\in\Gamma:\mathbf x+\mathbf e_\nu\in\Gamma\},
\qquad
\nu=1,\ldots,d.
\]
Thus $|\mathcal B_{\Gamma,\nu}|=(\ell-1)\ell^{d-1}$,
independently of $\nu$. This equality is why we consider cubes:
it allows us to center the local energy at a fixed fraction
of the total energy. Define the energy contained in $\Gamma$
and the fraction of lattice bonds contained in $\Gamma$ by
\[
H_\Gamma
=
-\sum_{\nu=1}^d
\sum_{\mathbf x\in\mathcal B_{\Gamma,\nu}}
\left(
c_{\mathbf x}^*c_{\mathbf x+\mathbf e_\nu}
+
c_{\mathbf x+\mathbf e_\nu}^*c_{\mathbf x}
\right),
\qquad
\nu_\Gamma
=
\frac{(\ell-1)\ell^{d-1}}{L^d}.
\]
\begin{theorem}\label{thm:local-energy}
For every fixed $d\geq1$ there exists $C_d<\infty$ such that, for every
side length $L$, every particle number $0<N<L^d$, every interval
${\cal I}\subset\ZZ/L\ZZ$ of length $1\leq\ell<L$, every energy $E$, and every
normalized $\varphi\in\mathcal H_N\cap\operatorname{Ran}\Pi_E$, 
one has
\[
\left\langle
\varphi,(H_\Gamma-\nu_\Gamma E)^2\varphi
\right\rangle
\leq
C_dNL^{d-1},
\qquad
\Gamma={\cal I}^d.
\]
Moreover, for every $\eta>0$,
\[
\left\|
\mathbf 1_{\{|H_\Gamma-\nu_\Gamma E|>N\eta\}}
\varphi
\right\|^2
\leq
\frac{C_dL^{d-1}}{N\eta^2}.
\]
\end{theorem}

\proof
Write $H_\Gamma=\mathrm d\Gamma(h_\Gamma)$. In the momentum basis,
\begin{equation}
\label{eq:local-energy-matrix-d}
(h_\Gamma)_{\mathbf k\mathbf q}
=
-\frac1V
\sum_{\nu=1}^d
\left(\e^{\i q_\nu}+\e^{-\i k_\nu}\right)
\sum_{\mathbf x\in\mathcal B_{\Gamma,\nu}}
\e^{\i(\mathbf q-\mathbf k)\cdot\mathbf x}.
\end{equation}
In particular, $(h_\Gamma)_{\mathbf k\mathbf k}=\nu_\Gamma\varepsilon(\mathbf k)$.  Set $g_\Gamma=h_\Gamma-\nu_\Gamma h_0$. 
Then $(g_\Gamma)_{\mathbf k\mathbf k}=0$. Moreover,
\[
\|h_\Gamma\|\leq2d,
\qquad
\|h_0\|\leq2d,
\qquad
\|g_\Gamma\|\leq4d,
\]
so that
\[
\sup_{\mathbf q\in K}
\sum_{\mathbf k\in K}
|(g_\Gamma)_{\mathbf k\mathbf q}|^2
\leq
16d^2.
\]
If $\mathbf r\neq0$, then
\[
|(g_\Gamma)_{\mathbf q+\mathbf r,\mathbf q}|^2
\leq
\frac{4d}{V^2}
\sum_{\nu=1}^d|\widehat\chi_{\mathcal B_{\Gamma,\nu}}(\mathbf r)|^2.
\]
By the counting argument in Lemma~\ref{lem:zero-transfer-hs},
for each $\mathbf r\neq0$ the equation
$\varepsilon(\mathbf q+\mathbf r)=\varepsilon(\mathbf q)$
has at most $2L^{d-1}$ solutions. Hence, by Parseval's identity,
\[
\|(g_\Gamma)_0\|_{\rm HS}^2
\leq
\frac{8dL^{d-1}}{V^2}
\sum_{\nu=1}^d
\sum_{\mathbf r\in K}|\widehat\chi_{\mathcal B_{\Gamma,\nu}}(\mathbf r)|^2
=
\frac{8dL^{d-1}}{V}
\sum_{\nu=1}^d|\mathcal B_{\Gamma,\nu}|
\leq
8d^2L^{d-1}.
\]
Proposition~\ref{prop:quadratic-general}, with
$\alpha=\nu_\Gamma$ and $\beta=0$, yields the result.
\qed

The particle number and energy in $\Gamma$ can also be
controlled simultaneously. Since every hopping term in
$H_\Gamma$ has both endpoints in $\Gamma$,
$[N_\Gamma,H_\Gamma]=0$.
Let $P^N_{{\rm eq},\eta}$ and $P^H_{{\rm eq},\eta}$ be the spectral
projections corresponding to
\[
\left|
N_\Gamma-\mu N
\right|
\leq
N \eta,
\qquad
\left|
H_\Gamma -\nu_\Gamma E
\right|
\leq
N \eta,
\]
respectively.

\begin{corollary}\label{thm:joint-concentration}
For every fixed $d\geq1$ there exists $C_d<\infty$ such that
\[
\left\|
\left(
I_{\mathcal H_N}
-
P^N_{{\rm eq},\eta}P^H_{{\rm eq},\eta}
\right)
\varphi
\right\|^2
\leq
\frac{C_dL^{d-1}}{N\eta^2}
\]
for every nonempty cube $\Gamma={\cal I}^d$ properly contained in $\Lambda_{L,d}$, every normalized
$\varphi\in\mathcal H_N\cap\operatorname{Ran}\Pi_E$, and every $\eta>0$.
\end{corollary}

\proof
The two projections commute, and
\[
I_{\mathcal H_N}
-
P^N_{{\rm eq},\eta}P^H_{{\rm eq},\eta}
\leq
\left(I_{\mathcal H_N}-P^N_{{\rm eq},\eta}\right)
+
\left(I_{\mathcal H_N}-P^H_{{\rm eq},\eta}\right).
\]
The claim follows from Theorems~\ref{thm:main} and
\ref{thm:local-energy}.
\qed

The same argument applies to any fixed finite family of commuting
quadratic observables satisfying the hypotheses of
Proposition~\ref{prop:quadratic-general}, with the constant in the
joint concentration bound equal to the sum of the individual constants.
\section{The One-Dimensional Refinement}
\label{sec-one-dimensional}
{\emergencystretch=1em
In this section $d=1$ and $\Gamma\subset\Lambda$ is an interval.
We use the structure of the one-particle energy eigenspaces
to prove Theorem~\ref{thm:one-dimensional-expectation}.\par}

\begin{lemma}\label{lem:one-shell}
Let $X,Y\subset K$, $|X|=|Y|=N$, satisfy $E(X)=E(Y)$. If
$\langle\Psi_X,N_\Gamma\Psi_Y\rangle\neq0$ and $X\neq Y$, then $X$ is
obtained from $Y$ by replacing one occupied momentum $q$ by $-q$.
\end{lemma}

\proof
By Lemma~\ref{lem:matrix-elements}, $X$ and $Y$ differ by one replacement,
say $q$ is replaced by $k$. Equality of the many-body energies gives
$\varepsilon(k)=\varepsilon(q)$. For
$\varepsilon(k)=-2\cos k$, this implies $k=q$ or $k=-q$ modulo $2\pi$.
Since $X\neq Y$, only the second alternative is possible.
\qed

\noindent{\bf Proof of Theorem~\ref{thm:one-dimensional-expectation}. } 
Let $\varphi\in\mathcal H_N\cap\operatorname{Ran}\Pi_E$ be normalized.
Only the zero-transfer part contributes to the expectation:
$\langle\varphi,B_\Gamma\varphi\rangle
=\langle\varphi,B_0\varphi\rangle$.
Since $b_{q,q}=0$, Lemma~\ref{lem:one-shell} gives
\[
B_0
=
\sum_{\substack{q\in K\\ q\neq -q}}
b_{-q,q}a_{-q}^*a_q.
\]

In the Slater basis, each term replaces an occupied momentum $q$ by $-q$.
As in the proof of Lemma~\ref{lem:matrix-elements}, the absolute row and
column sums of the matrix of $B_0$ are therefore bounded by
\[
\sup_{\substack{X\subset K\\ |X|=N}}
\sum_{q\in X}|b_{-q,q}|.
\]
The Schur test consequently yields
\begin{equation}
\label{eq:one-dimensional-schur}
\|B_0\|
\leq
\sup_{\substack{X\subset K\\ |X|=N}}
\sum_{q\in X}|b_{-q,q}|.
\end{equation}

Write $\Gamma=\{x_0,\ldots,x_0+\ell-1\}$. 
For $q\neq -q$,
\beq
b_{-q,q}
=
\frac1L\widehat\chi_\Gamma(2q)
=
\frac1L
\sum_{x=x_0}^{x_0+\ell-1}\e^{2\i qx}.
\label{mar-hot}
\eeq
Introduce the distance
\[
\delta_L(q)
=
\frac{L}{2\pi}
\operatorname{dist}
\bigl(q,\{0,\pi\}+2\pi\ZZ\bigr).
\]
For $q\neq-q$, \eqref{mar-hot} gives
\[
|b_{-q,q}|
\leq
\min\left\{1,\frac{1}{L|\sin q|}\right\}
\leq
\frac{C}{1+\delta_L(q)}.
\]
For $q=-q$, the coefficient vanishes.

For every integer $j\geq0$, at most four momenta $q\in K$
satisfy $j\leq\delta_L(q)<j+1$.
Hence, for every $N$-element set $X\subset K$,
\[
\sum_{q\in X}|b_{-q,q}|
\leq
C\sum_{j=1}^N\frac1j
\leq
C(1+\log N).
\]
Together with \eqref{eq:one-dimensional-schur}, this gives $\|B_0\|\leq C(1+\log N)$.  Finally, 
\[
\left|
\langle\varphi,N_\Gamma\varphi\rangle-N\mu
\right|
=
\left|
\langle\varphi,B_\Gamma\varphi\rangle
\right|
=
\left|
\langle\varphi,B_0\varphi\rangle
\right|
\leq
C(1+\log N).
\]
\qed
\appendix
\section{Proof of Mann's theorem}
\label{app:mann}

\noindent{\bf Proof of Theorem~\ref{thm:mann}. }
After a common rotation we may assume that $\zeta_1=1$. Let $n$ be the
smallest positive integer such that all $\zeta_j$ belong to $\mu_n$, the
multiplicative group of $n$th roots of unity. We prove that $n$ is
square-free and that every prime divisor of $n$ is at most $m$. This implies
$n\mid P_m$ and proves the theorem.

Fix a prime divisor $p$ of $n$ and write $p^a\Vert n$, meaning
that $p^a$ divides $n$ but $p^{a+1}$ does not. Set
$\rho=\e^{2\pi\i/p^a}$ and let
\[
K_0=\QQ(\mu_{n/p})=\QQ(\e^{2\pi\i p/n})
\]
be the field generated by the $(n/p)$th roots of unity.
 The
quotient group $\mu_n/\mu_{n/p}$ has order $p$, and the class of $\rho$
generates it. Consequently, every $\zeta_j\in\mu_n$ has a unique
representation
\[
\zeta_j=\xi_j\rho^{s_j},
\qquad
\xi_j\in\mu_{n/p},
\qquad
s_j\in\{0,\ldots,p-1\}.
\]
Grouping the original relation according to the exponent $s_j$ gives
\begin{equation}
\label{eq:mann-grouping}
\sum_{s=0}^{p-1}A_s\rho^s=0,
\qquad
A_s=
\sum_{j:s_j=s}\xi_j\in K_0.
\end{equation}
Suppose first that $a\geq2$. The representation
$\zeta=\xi\rho^s$, with $\xi\in\mu_{n/p}$, shows that
$K_0(\rho)=\QQ(\mu_n)$. 
The standard cyclotomic degree formula gives
\footnote{Here $\varphi(r)$ denotes Euler's totient function, the number of
integers in $\{1,\ldots,r\}$ relatively prime to $r$. We use the standard
identity $[\QQ(\mu_r):\QQ]=\varphi(r)$.}
\[
[K_0(\rho):K_0]
=
\frac{\varphi(n)}{\varphi(n/p)}
=p.
\]
Moreover, $\rho^p\in\mu_{n/p}\subset K_0$, so $\rho$ is a root of
$X^p-\rho^p\in K_0[X]$. Since the extension has degree $p$, this polynomial
is the minimal polynomial of $\rho$ over $K_0$. Hence
$1,\rho,\ldots,\rho^{p-1}$ are linearly independent over $K_0$, and
\eqref{eq:mann-grouping} implies
\[
A_0=\cdots=A_{p-1}=0.
\]
The class $s=0$ is nonempty because $\zeta_1=1$. It is also proper: if all
$s_j$ were zero, all roots would belong to $\mu_{n/p}$, contradicting the
minimality of $n$. Thus $A_0=0$ is a proper vanishing subsum of the original
relation, contrary to its minimality. We conclude that $a=1$. Since $p$ was
arbitrary, $n$ is square-free.

Now let $p$ be an odd prime divisor of $n$. Since $n$ is square-free,
\[
[K_0(\rho):K_0]
=
\frac{\varphi(n)}{\varphi(n/p)}
=p-1.
\]
The primitive $p$th root $\rho$ satisfies
\[
1+\rho+\cdots+\rho^{p-1}=0,
\]
and the polynomial $1+X+\cdots+X^{p-1}$ has degree $p-1$. It is therefore
the minimal polynomial of $\rho$ over $K_0$. The polynomial
$\sum_{s=0}^{p-1}A_sX^s$ in \eqref{eq:mann-grouping} has degree at most
$p-1$ and vanishes at $\rho$, so it is a scalar multiple of the minimal
polynomial. Hence $A_0=A_1=\cdots=A_{p-1}$. 
If $p>m$, at least one of the $p$ residue classes is empty, because the
relation contains only $m$ terms. The corresponding $A_s$ is zero, and the
preceding equality then forces all $A_s$ to vanish. As above, $A_0=0$ gives
a proper vanishing subsum, a contradiction. Thus $p\leq m$. The prime $2$
is automatically at most $m$, since a nontrivial vanishing relation has at
least two terms. Therefore every prime divisor of $n$ is at most $m$, and
$n\mid P_m$. \qed

\bibliographystyle{cap-alpha}
\bibliography{MASTER}

\end{document}